\documentclass[%
 reprint,
 amsmath,amssymb,
 aps,
]{revtex4-2}
\usepackage{graphicx}% Include figure files
\usepackage{dcolumn}% Align table columns on decimal point
\usepackage{bm}% bold math
\usepackage{amsmath}
\usepackage{empheq}
\usepackage{tikz}
\usetikzlibrary{graphs,quotes}
\usetikzlibrary{math}
\usetikzlibrary{calc}
\begin{document}

%\preprint{APS/123-QED}

\title{Bounds on the rates of statistical divergences and mutual information
via stochastic thermodynamics}
% Force line breaks with \\

\author{Jan Karbowski$^{*}$}

%\begin{center}
\affiliation{\it  Institute of Applied Mathematics and Mechanics,
Department of Mathematics, Informatics, and Mechanics,
University of Warsaw, ul. Banacha 2, 02-097 Warsaw, Poland}
%\end{center}


%\author{Author Two} 
%\affiliation{Affiliation2}%


\begin{abstract}
Statistical divergences are important tools in data analysis, information theory,
and statistical physics, and there exist well known inequalities on their bounds.
However, in many circumstances involving temporal evolution, one needs limitations
on the rates of such quantities, instead. Here, several general upper bounds on
the rates of some f-divergences are derived, valid for any type of stochastic
dynamics (both Markovian and non-Markovian), in terms of information-like
and/or thermodynamic observables. As special cases, the analytical bounds on
the rate of mutual information are obtained. The major role in all those limitations
is played by temporal Fisher information, characterizing the speed of global system
dynamics, and some of them contain entropy production, suggesting a link with
stochastic thermodynamics. Indeed, the derived inequalities can be used for
estimation of minimal dissipation and global speed in thermodynamic stochastic systems.
Specific applications of these inequalities in physics and neuroscience are given,
which include the bounds on the rates of free energy and work in nonequilibrium systems,
limits on the speed of information gain in learning synapses, as well as the bounds on
the speed of predictive inference and learning rate. Overall, the derived bounds can
be applied to any complex network of interacting elements, where predictability and
thermodynamics of network dynamics are of prime concern.
  
\begin{description}
\item[Keywords]
  rate of f-divergence, bounds, Tsallis and Renyi divergences,
  rate of mutual  \\
  information, stochastic thermodynamics, information thermodynamics.
\end{description}
\end{abstract}


\maketitle

%\tableofcontents

\section{\label{sec:level1}Introduction}

Statistical divergences, or distances, known as f-divergences are commonly
used to quantify how much different are two probability distributions
\cite{renyi,csiszar}. The most popular special cases of these divergences
are: Renyi divergence \cite{renyi}, Tsallis divergence \cite{tsallis1}, and
Kullback-Leibler (KL) divergence \cite{kullback}, which is a limiting case of
the former two (for a review see: \cite{cichocki}). In statistical physics,
KL and Tsallis divergences have prominent roles and have been shown to
relate to information gain and other important physical quantities, like entropy
production, work, and other observables \cite{hatano,esposito,still,falasco}.
In computer science, and recently in machine learning, KL has been used, among other
things, in assessing coding accuracy and efficiency \cite{cover,reid}. Moreover,
f-divergences have many applications in classic information theory \cite{liese},
and in the emerging field of information geometry \cite{amari}. There are many
inequalities relating different types of divergences and inequalities bounding
them from above \cite{csiszar,sason}. However, virtually all of these relations
and bounds apply only to static (stationary) situations. Since physical quantities
generally depend on time, probability distributions describing them are often time
dependent. Consequently, in real physical systems, statistical divergences can also
change in time, and their variability can provide an important information about
predictability of probabilistic system's dynamics. It is known that for isolated
stochastic systems with Markov dynamics (either of master equation or Fokker-Planck
equation types) all f-divergences decrease monotonically with time between
time-dependent state probability distribution and its equilibrium distribution
\cite{renyi,morimoto,risken,borland,yamano2,gorban}, which can be interpreted as
a loss of information in autonomously relaxing systems \cite{cover}. However,
no such simple relation exists for statistical divergences between two arbitrary
time dependent distributions.

The goal of this paper is to shed some light on such more general conditions by
determining fundamental bounds on the rates of popular f-divergences for
arbitrary probability distributions, and relate these bounds to known observables.
The obtained bounds may have practical applications, as it is often difficult to
calculate exactly the rates of statistical divergences for a system at hand.
Moreover, and more importantly, such limits may have conceptual meaning, especially
in the connection with stochastic and information thermodynamics
(e.g. see \cite{still,falasco,crooks,parrondo,vu}), when we interpret stochastic
f-divergences as generalized information gains \cite{cover}.
Indeed, the inequalities found here for the rates of statistical divergences have
similar flavor to several inequalities discovered recently in stochastic thermodynamics
linking physical observables with information, entropy production, and the speed of
global dynamics \cite{shiraishi,ito,nicholson,koyuk,vo}.

In this work, two types of bounds on the rates of Tsallis, Renyi, and
Kullback-Leibler divergences are derived. The first type is very general
and consists of two inequalities related solely to kinematic characteristics.
The second type is more restrictive, as it applies only to Markov dynamics for
probabilities obeying master equation, and it consists of four inequalities
involving both kinematic and thermodynamic observables. These results are then
used to derive general bounds on the rate of mutual information between two
stochastic variables. As an example, a driven one-step Markov process is used to
illustrate the sharpness and ranking of the obtained bounds. These bounds can
be used naturally for determining lower speed limits on stochastic dynamics
and minimal dissipation. Various other, more specific, applications are also
presented, ranging from physics to neuroscience. These applications involve the
limits on the rates of free energy in nonequilibrium thermodynamic systems, as
well as the bounds on the speed of information gain, predictive inference, and
learning rate in neural systems.

\section{\label{sec:level2}Preliminaries}

\subsection{\label{sec:levell2} Statistical divergences and useful relations}

Consider a physical system that has internal states labeled by index $n$, and
which can be described by two time dependent probability distributions $p_{n}(t)$
and $q_{n}(t)$. Although it is not essential for the arguments below, it is
convenient to think about $q_{n}(t)$ as a true (reference) probability distribution
of the system stochastic dynamics, and about $p_{n}(t)$ as its estimation or prediction.
Before we introduce f-divergences, let us first define a helpful quantity, which
can be called $\alpha$-coefficient $C_{\alpha}(p||q)$ between the distributions
$p$ and $q$ (also known as Chernoff $\alpha$-coefficient or divergence;
\cite{kolchinsky}):

\begin{eqnarray}
  C_{\alpha}(p||q)= \langle \Big(\frac{p}{q}\Big)^{\alpha-1}\rangle_{p}, 
\end{eqnarray}
where $\alpha$ is arbitrary real number, and the symbol
$\langle \big(p/q\big)^{\alpha-1}\rangle_{p} =
\sum_{n} p_{n}(t)\big(\frac{p_{n}(t)}{q_{n}(t)}\big)^{\alpha-1}$,
which means averaging with respect to probability distribution $p$.
Note that equivalently $C_{\alpha}= \langle (p/q)^{\alpha}\rangle_{q}$,
which implies $C_{\alpha}(p||p)= 1$ for all $\alpha$, and also
$C_{0}(p||q)= C_{1}(p||q)= 1$. (The focus is on discrete states but the
results are also valid for continuous variables through replacing sums
by integrals, and such transformations are done below occasionally.)
The $\alpha$-coefficient provides a core for basic f-divergences,
and thus can be of interest in itself.

Two major f-divergences, Tsallis $T_{\alpha}$ and Renyi $R_{\alpha}$, between
$p$ and $q$ distributions are expressed in terms of $\alpha$-coefficient
as \cite{cichocki}

%\begin{widetext}
\begin{eqnarray}
  T_{\alpha}(p||q)= \frac{C_{\alpha}(p||q) - 1}
  {\alpha-1},
\end{eqnarray}
%\end{widetext}
and
\begin{eqnarray}
  R_{\alpha}(p||q)= \frac{\ln C_{\alpha}(p||q)}
  {\alpha-1},
\end{eqnarray}
which implies a simple relationship between them as
$R_{\alpha}= \ln\big(1+(\alpha-1)T_{\alpha}\big)/(\alpha-1)$.
When $T_{\alpha}(p||q)$ and $R_{\alpha}(p||q)$ are close to 0,
then the probability distribution $p$ approximates
or predicts the distribution $q$ very well.

In the limit $\alpha\mapsto 1$, Tsallis and Renyi divergences both tends
to KL divergence $D_{KL}$, known also as relative entropy \cite{cover},
i.e., $T_{1}= R_{1}= D_{KL}(p||q)= \sum_{n} p_{n}(t)\ln\frac{p_{n}(t)}{q_{n}(t)}$.
For $\alpha=1/2$, we obtain $T_{1/2}$ as a Hellinger distance, while for
$\alpha=2$ our $T_{2}$ is the Pearson $\chi^{2}$-divergence.

In our derivations, we will also need two inequalities. The first is the stochastic
version of the generalized H\"older inequality, which for $m$ arbitrary stochastic
variables $X_{1}, X_{2},..., X_{m}$ takes the form \cite{finner}

\begin{eqnarray}
  \langle |X_{1}...X_{m}|\rangle
  \le \langle |X_{1}|^{1/\lambda_{1}}\rangle^{\lambda_{1}}
  ... \langle |X_{m}|^{1/\lambda_{m}}\rangle^{\lambda_{m}},
%\nonumber  \\
\end{eqnarray}
where $\lambda_{i}$ are positive real numbers such that $\sum_{i=1}^{m} \lambda_{i}= 1$,
the symbol $|...|$ means the absolute value, and $\langle ...\rangle$ denotes
averaging with respect to some probability distribution. The equality in Eq. (4)
is achieved when there are proportionalities between all the rescaled variables,
i.e., when $|X_{i}|^{1/\lambda_{i}} = c_{i}|X_{1}|^{1/\lambda_{1}}$ for every $i=2,...,m$,
where $c_{i}$ are some positive (possibly time dependent) coefficients. When $m=2$
and $\lambda_{1}= \lambda_{2}= 1/2$, Eq. (4) becomes a classic Cauchy-Schwartz
inequality.

The second useful inequality relates arithmetic and geometric means to the
so-called logarithmic mean of two positive numbers $x$ and $y$, and reads \cite{carlson}:

\begin{eqnarray}
\sqrt{xy}  \le \frac{x-y}{\ln(x)-\ln(y)}  \le  \frac{x+y}{2}.
\end{eqnarray}

With Eqs. (1-5), we have all necessary ingredients to derive the upper bounds on
the rates of f-divergences.

\subsection{\label{sec:levell3} Rates of statistical divergences.}

Because of the relations (1-3), the rates of Tsallis and Renyi divergences
can be expressed in terms of the rate of $\alpha$-coefficient. For that reason,
and because calculations are a little easier for $C_{\alpha}$, below we focus
on the temporal rate of $C_{\alpha}$ and its bounds. The bounds for the
rates of $T_{\alpha}$ and $R_{\alpha}$ are obtained as straightforward extensions
of the bounds on $dC_{\alpha}/dt$.

The temporal rate of $C_{\alpha}(p||q)$ can be written
as

\begin{eqnarray}
  \frac{dC _{\alpha}}{dt}=
  \alpha \sum_{n} \dot{p}_{n} \big[\big(p_{n}/q_{n}\big)^{\alpha-1} - C_{\alpha}\big]
  \nonumber  \\   
  - (\alpha-1) \sum_{n} \dot{q}_{n} \big[\big(p_{n}/q_{n}\big)^{\alpha} - C_{\alpha}\big]
  \nonumber  \\
  \le     |\frac{dC _{\alpha}}{dt}|  \le 
  |\alpha|  \langle |\frac{\dot{p}}{p}\big[\big(p/q\big)^{\alpha-1}
    - C_{\alpha}\big]| \rangle_{p} \nonumber  \\   
  + |\alpha-1| \langle |\frac{\dot{q}}{q} \big[\big(p/q\big)^{\alpha}
   - C_{\alpha}\big] | \rangle_{q} 
\end{eqnarray}         
where $\dot{p}_{n}= dp_{n}/dt$ and similar for $\dot{q}_{n}$. The notation
$\langle ...\rangle_{p}$ means averaging with respect to distribution $p_{n}$.
In both bracket terms we subtracted $C_{\alpha}$ for convenience (see below),
but this trick does not change the result of summation, as
$\sum_{n} \dot{p}_{n}= \sum_{n} \dot{q}_{n}= 0$.
In the last inequality we used a well known relation $x+y \le |x+y| \le |x|+|y|$,
for arbitrary real numbers $x,y$.

Eq. (6) enables us to write the upper limit on the rate of Kullback-Leibler
divergence $D_{KL}(p||q)$, since $D_{KL}= \lim_{\alpha\mapsto 1} (C_{\alpha}-1)/(\alpha-1)$,
and $D_{KL}= T_{1}$. We have

\begin{eqnarray}
 \frac{dD_{KL}}{dt}
 \le  \langle |\frac{\dot{p}}{p}|
 |\ln\big(\frac{p}{q}\big) - D_{KL}| \rangle_{p}
  +  \langle |\frac{\dot{q}}{q}|
   |\big(\frac{p}{q}\big) - 1| \rangle_{q},  
   \nonumber  \\
 \frac{dD_{KL}}{dt}
 \le  \langle |\dot{p}/p|
 |\ln(p/q) - D_{KL}| \rangle_{p}
  + \langle |\frac{d\ln(p/q)}{dt}| \rangle_{p},     
   \nonumber  \\
\end{eqnarray}         
where we used the fact that
$\lim_{\alpha\mapsto 1} \big[(p/q)^{\alpha-1}-1\big]/(\alpha-1) =
\ln(p/q)$.

Our goal in the next sections is to find upper bounds on
$dC_{\alpha}/dt$ and $dD_{KL}/dt$, which in effect is equivalent
to determining the limits on the averages in Eqs. (6) and (7).
Having the bounds on $dC_{\alpha}/dt$, it is easy to obtain the
upper limits on the rates of Tsallis and Renyi divergences, since
from Eqs. (2) and (3) it follows that
$dC_{\alpha}/dt= (\alpha-1)dT_{\alpha}/dt$ 
and
$dC_{\alpha}/dt= (\alpha-1)e^{(\alpha-1)R_{\alpha}} dR_{\alpha}/dt$.

\subsection{\label{sec:levell3} Temporal and relative Fisher information.}

In the derivation below it will be used the so-called temporal Fisher information,
which is defined for probability distribution $p_{n}$ as \cite{frieden,ito}

\begin{eqnarray}
I_{F}(p)= \sum_{n} p_{n} \Big(\frac{\dot{p}_{n}}{p_{n}}\Big)^{2}
 \equiv  \langle \big(\dot{p}/p\big)^{2}\rangle_{p},
\end{eqnarray}         
and analogically for the distribution $q_{n}$. Note that the role of the
control/external parameter is played here by the time. The quantity $I_{F}(p)$
is usually interpreted as a square of the speed of global system dynamics
described by the distribution $p_{n}$ \cite{frieden,ito}. For example, if
$p_{n}$ is a Poisson distribution with time-dependent intensity parameter
$\nu$, i.e., $p_{n}= (\nu^{n}/n!)e^{-\nu}$, then the temporal Fisher information
is $I_{F}(p)= (\dot{\nu})^{2}/\nu$, where $\dot{\nu}$ is temporal derivative
of $\nu$. This result indicates that $\sqrt{I_{F}(p)}$ is proportional to the
absolute speed of changes in the intensity parameter.

By a direct extension, we can define a relative temporal Fisher information
$F(p||q)$ between two probability distributions $p_{n}$ and $q_{n}$ as

\begin{eqnarray}
F(p||q)= \langle \Big(\frac{d\ln(p/q)}{dt}\Big)^{2}\rangle_{p}
 \equiv  \langle \big(\dot{p}/p - \dot{q}/q\big)^{2}\rangle_{p}.
\end{eqnarray}
This definition, to the author's knowledge, is the first definition of
the relative temporal Fisher information, where time is the control
parameter. Definition (9) is the generalization of a more standard 
relative Fisher information with non-temporal control parameter
\cite{otto,yamano1,yamano2}, which however has not received much attention
in physics. $F(p||q)$ can be interpreted as a measure of relative speeds
of system dynamics described by two different distributions $p_{n}$ and
$q_{n}$. Additionally, $F(p||q)= 0$, if and only if $p_{n}(t)= q_{n}(t)$
for all $n$. To get more intuitive understanding of $F(p||q)$ let us
take again Poisson distribution for $p$ and $q$ with time dependent
intensity parameters $\nu_{1}$ and $\nu_{2}$, respectively. Then it can be
shown that
$F(p||q)= \nu_{1}[(\dot{\nu}_{1}/\nu_{1})
- (\dot{\nu}_{2}/\nu_{2})]^{2}
+ (\dot{\nu}_{2}/\nu_{2})^{2}(\nu_{1}-\nu_{2})^{2}$.
This means that $F(p||q)$ is zero at any given time only when both intensity
parameters and their speeds are equal.

\section{\label{sec:level3} General kinematic bounds on the rates of divergences}  

%\subsection{\label{sec:levell3} Kinematic bound on the rate of divergences.}

In this section we derive upper bounds on the rates of $T_{\alpha}$, $R_{\alpha}$,
and $D_{KL}$, which we call the kinematic bounds.

\subsection{\label{sec:levell3} Limits on rates of divergences via Fisher information.}

\subsubsection{ Rates of $\alpha$-coefficient and Tsallis and Renyi divergences.}

Application of Eq. (4) for $m=2$ and $\lambda_{1}=\lambda_{2}=1/2$, with
$X_{1}= \frac{\dot{p}}{p}$ and
$X_{2}= \big[\big(\frac{p}{q}\big)^{\alpha-1} - C_{\alpha}\big]$
for the first term on the right in the last line of Eq. (6), and similarly for
the second term in that line yields

\begin{eqnarray}
 | \frac{dC_{\alpha}}{dt}| \le  
  |\alpha| \sqrt{\langle \Big(\frac{\dot{p}}{p}\Big)^{2}\rangle_{p}}
 \sqrt{\langle \big[(p/q)^{\alpha-1} - C_{\alpha}\big]^{2}\rangle_{p}}
  \nonumber  \\   
+ |\alpha-1| \sqrt{\langle \Big(\frac{\dot{q}}{q}\Big)^{2}\rangle_{q}}
 \sqrt{\langle \big[(p/q)^{\alpha} - C_{\alpha}\big]^{2} \rangle_{q}}.  
\end{eqnarray}         
The ratios $\langle\big(\dot{p}/p\big)^{2}\rangle_{p}$, and
$\langle\big(\dot{q}/q\big)^{2}\rangle_{q}$ can be identified with temporal Fisher
informations as in Eq. (8). The final step is to note that

\begin{eqnarray}
%\begin{subequations}
%\begin{align}
%\text{Level 1}&:
  \langle \big[(p/q)^{\alpha-1} - C_{\alpha}\big]^{2}\rangle_{p}
=  C_{2\alpha-1} - C_{\alpha}^{2},
  \nonumber  \\   
%\end{eqnarray*}         
%and
%\begin{eqnarray*}
%\text{Level 2}&:
\langle \big[(p/q)^{\alpha} - C_{\alpha}\big]^{2}\rangle_{q}
= C_{2\alpha}-C_{\alpha}^{2}.
%\end{align}
%\end{subequations}
\end{eqnarray}
Interestingly, the above averages correspond to variances of
$(p_{n}/q_{n})^{\alpha-1}$ and $(p_{n}/q_{n})^{\alpha}$ around
$C_{\alpha}$ averaged with respect to either $p_{n}$ or $q_{n}$ distributions.

After these substitutions, the general upper bound on $dC_{\alpha}/dt$
is given by

\begin{eqnarray}
 | \frac{dC_{\alpha}}{dt}| \le
  |\alpha|\sqrt{I_{F}(p)}\sqrt{C_{2\alpha-1}-C_{\alpha}^{2}}
  \nonumber  \\
  +  |\alpha-1|\sqrt{I_{F}(q)}\sqrt{C_{2\alpha}-C_{\alpha}^{2}}.
\end{eqnarray}
The right hand side of this inequality is a kinematic limit set on
the dynamics of $\alpha$-coefficient, and it is called the bound B1
hereafter.

After using transformations in Eqs. (2) and (3), the inequality (12)
allows us to find the upper bounds on the rates of Tsallis and Renyi
divergences, $dT_{\alpha}/dt$ and $dR_{\alpha}/dt$. They are given by

\begin{eqnarray}
 |\frac{dT_{\alpha}}{dt}| \le
 |\alpha| \sqrt{I_{F}(p)}\sqrt{\frac{2}{(\alpha-1)}(T_{2\alpha-1}-T_{\alpha})-T_{\alpha}^{2}}
  \nonumber  \\
  +  \sqrt{I_{F}(q)}\sqrt{(2\alpha-1)T_{2\alpha}-2(\alpha-1)T_{\alpha}
    - (\alpha-1)^{2}T_{\alpha}^{2}},
  \nonumber  \\
\end{eqnarray}
and
\begin{eqnarray}
  | \frac{dR_{\alpha}}{dt}| \le \frac{|\alpha|}{|\alpha-1|}
    \sqrt{I_{F}(p)}\sqrt{e^{2(\alpha-1)(R_{2\alpha-1}-R_{\alpha})} - 1}
  \nonumber  \\
  +  \sqrt{I_{F}(q)}\sqrt{e^{[(2\alpha-1)R_{2\alpha}-2(\alpha-1)R_{\alpha}]} - 1}.
\end{eqnarray}
Eqs. (12-14) constitute the first major result of this paper. They imply
that the temporal rates of the Tsallis and Renyi divergence are bounded by the
products of the global rates of system dynamics and various nonlinear combinations
of associated divergences. It is good to keep in mind that inequalities (12-14)
have a general character that is independent on the nature of dynamics of
probabilities, i.e., valid for both Markovian and non-Markovian dynamics.
Moreover, the core Eq. (12) is structurally similar to the upper bound on the
average rate of a stochastic observable, recently investigated \cite{ito,nicholson}.

In a special case when the distribution $q$ is the steady-state distribution
of $p$, i.e., $q= p_{\infty}$, we have temporal Fisher information $I_{F}(q)=0$,
and Eqs. (13) and (14) simplify considerably. For instance, for $\alpha= 2$,
from Eq. (13) we obtain the bound on the rate of Pearson divergence as
$|dT_{2}/dt| \le 2\sqrt{I_{F}(p)}\sqrt{2(T_{3}-T_{2})-T_{2}^{2}}$.

\subsubsection{An example: Weakly time dependent exponential distributions.}

In order to gain an insight about the limit set on the rate of
$\alpha$-coefficient in Eq. (12), it is instructive to analyze an explicit
example. Consider two time dependent continuous distributions
$p_{x}(x)= \nu_{1} e^{-\nu_{1} x}$ and $q_{x}(x)= \nu_{2} e^{-\nu_{2} x}$, where
$\nu_{1}$ and $\nu_{2}$ are time dependent but positive. Furthermore, let
us assume that $\nu_{1}(t)= \nu + \epsilon(\Delta + r_{1}(t))$, and
$\nu_{2}(t)= \nu + \epsilon r_{2}(t)$, where the parameter $\epsilon \ll 1$,
$\nu$ and $\Delta$ are time independent, and $r_{1}(t)$ and $r_{2}(t)$
are arbitrary time dependent functions. This parameterization means
that $\nu_{1}(t)$ and $\nu_{2}(t)$, and their rates, differ only slightly
($\sim \epsilon$) at any moment of time. For this example, we can find
explicit simple expressions for all quantities in Eq. (12) to the lowest order
in $\epsilon$ (see Appendix A). From this it follows that the left hand side
of Eq. (12) is equal to
$|\alpha(\alpha-1)|(\epsilon/\nu)^{2}
|r_{1}-r_{2}+\Delta||\dot{r}_{1}-\dot{r}_{2}| + O(\epsilon^{3})$, and
the right hand side is
$|\alpha(\alpha-1)|(\epsilon/\nu)^{2}
|r_{1}-r_{2}+\Delta|(|\dot{r}_{1}|+|\dot{r}_{2}|) + O(\epsilon^{3})$,
where $\dot{r}_{1}, \dot{r}_{2}$ are time derivatives of $r_{1}, r_{2}$.
Thus the difference between the two sides is set only by the difference
between $|\dot{r}_{1}-\dot{r}_{2}|$ and  $(|\dot{r}_{1}|+|\dot{r}_{2}|)$,
i.e. by the relative speeds of $r_{1}$ and $r_{2}$. The equality of both
sides in Eq. (12) is achieved at the moments when $\dot{r}_{1}$ and
$\dot{r}_{2}$ have the opposite signs, irrespective of the functional
dependence of $r_{1}(t)$ and $r_{2}(t)$.

%\subsection{\label{sec:levell3} Rate of Kullback-Leibler divergence.}
\subsubsection{Rate of Kullback-Leibler divergence.}

From the first line of Eq. (7) and the above considerations it immediately
follows the upper bound on the rate $dD_{KL}/dt$:

\begin{eqnarray}
 | \frac{dD_{KL}}{dt}| \le
  \sqrt{I_{F}(p)}\sqrt{\langle \ln^{2}(p/q)\rangle_{p} - D_{KL}^{2}}
  \nonumber  \\
  +  \sqrt{I_{F}(q)}\sqrt{T_{2}}.
\end{eqnarray}
The bound (15) involves the mean of the logarithm square, i.e.,
$\langle \ln^{2}(p/q)\rangle_{p}$, which may be difficult to compute in
many practical situations. Therefore it is good to have an upper limit on
the logarithm in terms of other divergences. Such a limit is provided by
Eq. (5) (see also Eq. (C4)), which implies for probabilities $p$ and $q$:

\begin{eqnarray}
  \langle \ln^{2}(p/q)\rangle_{p} \le
  \langle \Big(\sqrt{\frac{p}{q}} - \sqrt{\frac{q}{p}}\Big)^{2}\rangle_{p}
  =  \langle \frac{p}{q}\rangle_{p} - 1 = T_{2}. 
  \nonumber  \\
\end{eqnarray}
Consequently, $dD_{KL}/dt$ is restricted also by

\begin{eqnarray}
 | \frac{dD_{KL}}{dt}| \le
  \sqrt{I_{F}(p)}\sqrt{T_{2} - D_{KL}^{2}}
%  \nonumber  \\
  +  \sqrt{I_{F}(q)}\sqrt{T_{2}}.
\end{eqnarray}
Obviously, the limit set by Eq. (15) is tighter than the one present in Eq. (17).
Moreover, the prominent role in the bound (17) is played by Pearson divergence
$T_{2}$.

\subsection{\label{sec:levell3} Limits on the rates of divergences via
  relative Fisher information.}

The rate $dC_{\alpha}/dt$ in Eq. (6) can be equivalently expressed as

\begin{eqnarray}
 \frac{dC_{\alpha}}{dt} =
\langle \frac{\dot{q}}{q} \Big[\Big(\frac{p}{q}\Big)^{\alpha} - C_{\alpha}\Big] \rangle_{q}
+ \alpha\langle \Big(\frac{p}{q}\Big)^{\alpha-1} \big[\dot{p}/p -\dot{q}/q\big] \rangle_{p}
\nonumber  \\
\le   \sqrt{I_{F}(q)} \sqrt{C_{2\alpha}- C_{\alpha}^{2}}   
+ |\alpha|\sqrt{F(p||q)} \sqrt{C_{2\alpha-1}},
\nonumber  \\
\end{eqnarray}         
where we used the Cauchy-Schwartz inequality and the definition (9) for 
relative temporal Fisher information. The last line of Eq. (18) is
another kinematic limit set on the dynamics of $\alpha$-coefficient,
and it is called the bound B2 below.

Eq. (18) gives us the upper bounds on
the rate of Tsallis divergence:

\begin{eqnarray}
 | \frac{dT_{\alpha}}{dt}| \le
  \sqrt{I_{F}(q)} \sqrt{\frac{(2\alpha-1)T_{2\alpha}-2(\alpha-1)T_{\alpha}}
    {(\alpha-1)^{2}} - T_{\alpha}^{2}}
\nonumber  \\
+ \frac{|\alpha|}{|\alpha-1|}\sqrt{F(p||q)} \sqrt{2(\alpha-1)T_{2\alpha-1}+1},
\nonumber  \\
\end{eqnarray}         
and on the rate of Renyi divergence: 

\begin{eqnarray}
 | \frac{dR_{\alpha}}{dt}| \le \frac{\sqrt{I_{F}(q)}}{|\alpha-1|}
  \sqrt{e^{(2\alpha-1)R_{2\alpha}-2(\alpha-1)R_{\alpha}} - 1}
\nonumber  \\
+ \frac{|\alpha|}{|\alpha-1|}\sqrt{F(p||q)} e^{(\alpha-1)(R_{2\alpha-1}-R_{\alpha})}.
\nonumber  \\
\end{eqnarray}

By the same token, from the second line of Eq. (7), the rate of Kullback-Leibler
divergence is limited by

\begin{eqnarray}
 | \frac{dD_{KL}}{dt}| \le
  \sqrt{I_{F}(p)} \sqrt{\langle\ln^{2}(p/q)\rangle - D_{KL}^{2}}   
+ \sqrt{F(p||q)}
\nonumber  \\
\le  \sqrt{I_{F}(p)} \sqrt{T_{2} - D_{KL}^{2}}   
+ \sqrt{F(p||q)},
\nonumber  \\
\end{eqnarray}         
where for the second term on the right we used Eq. (4) for $m=2$ with $X_{1}=1$,
$X_{2}= \frac{d\ln(p/q)}{dt}$, and $\lambda_{1}=\lambda_{2}= 1/2$.

Note that the rates $dT_{\alpha}/dt$, $dR_{\alpha}/dt$, $dD_{KL}/dt$ and
their bounds in Eqs. (13), (14), (15), (17) and (19-21) are all zero if
$p_{n}(t)= q_{n}(t)$ for all $n$,
regardless of the temporal dependence of these probabilities. This is because
of the properties: $T_{\alpha}(p||p)= R_{\alpha}(p||p)= F(p||p)= 0$ for all
$\alpha$.

\section{\label{sec:level3} Kinematic-thermodynamic bounds for Markov processes.}

In this section we derive bounds on $dC_{\alpha}/dt$, $dT_{\alpha}/dt$,
$dR_{\alpha}/dt$, and $dD_{KL}/dt$ that involve both kinematic and thermodynamic
variables. To do this we need to assume that the dynamics of both probability
distributions, $p_{n}$ and $q_{n}$, are Markovian and represented
by master equations \cite{schnakenberg}

\begin{eqnarray}
  \dot{p}_{n}= \sum_{k} (w_{nk}p_{k}-w_{kn}p_{n}),
  \nonumber  \\
  \dot{q}_{n}= \sum_{k} (v_{nk}q_{k}-v_{kn}q_{n}),  
\end{eqnarray}         
where $w_{kn}$ and $v_{kn}$ are corresponding transition rates for jumps from
$n$ to $k$ states. These aggregate transition rates can be composed of several
sub-transitions corresponding to distinct underlying physical processes labeled
by $s$, i.e. $w_{kn}= \sum_{s} w_{kn}^{(s)}$, and $v_{kn}= \sum_{s} v_{kn}^{(s)}$
\cite{esposito}.

\subsection{\label{sec:levell3} Bounds of first kind.}

\subsubsection{Rate of $\alpha$-coefficient.}

Consider the term
$\langle |\dot{p}/p||(p/q)^{\alpha-1} - C_{\alpha}|\rangle_{p}$
in the last two lines of Eq. (6).
The same steps can be taken for the other term in that equation, and hence
are omitted here.
First, we make the following decomposition:

%\begin{eqnarray*}
$\langle |\dot{p}/p| |(p/q)^{\alpha-1} - C_{\alpha}|\rangle_{p}=
\langle |\dot{p}/p|^{1/3} |\dot{p}/p|^{2/3}
|(p/q)^{\alpha-1} - C_{\alpha}|\rangle_{p}$. 
%\end{eqnarray*}         
Second, we apply the H\"older inequality (4) in the latter average with
$X_{1}= |\dot{p}/p|^{1/3}$, $X_{2}= |\dot{p}/p|^{2/3}$, 
and $X_{3}= |(p/q)^{\alpha-1} - C_{\alpha}|$, and for 
$\lambda_{1}= \lambda_{2}= \lambda_{3}= 1/3$. As a result, we obtain

\begin{eqnarray}
  \langle |\dot{p}/p|
  |\big(p/q\big)^{\alpha-1} - C_{\alpha}| \rangle_{p}
  \nonumber  \\
\le \Big( I_{F}(p) \langle|\dot{p}/p|\rangle_{p}
  \langle |\big(p/q\big)^{\alpha-1} - C_{\alpha}|^{3}\rangle_{p}\Big)^{1/3}.
\end{eqnarray}

The term $\langle|\dot{p}/p|\rangle_{p}$ can be  limited in two different
ways (see Appendix B). One way involves internal activity $A_{p}$ in the system
described by distribution $p$

\begin{eqnarray}
  \langle |\dot{p}/p| \rangle_{p} \le 2A_{p},
\end{eqnarray}
where $A_{p}$, which also can be called global average escape rate
is defined as \cite{baiesi,koyuk,vo}

\begin{eqnarray}
  A_{p} =  \frac{1}{2}\sum_{nk} (w_{nk}p_{k} + w_{kn}p_{n})
  \equiv \sum_{n}\overline{w}_{n}p_{n} = \langle\overline{w}\rangle_{p},
\end{eqnarray}         
where $\overline{w}_{n}= \sum_{k} w_{kn}$ is the total escape rate from state $n$.

The second way of bounding $\langle|\dot{p}/p|\rangle_{p}$ is through
(see Appendix B):

\begin{eqnarray}
  \langle |\dot{p}/p| \rangle_{p} \le
  \sqrt{2\dot{S}_{p}}\sqrt{A_{p}},
\end{eqnarray}
where $\dot{S}_{p}$ is the coarse-grained entropy production rate in the
system with the distriburtion $p$ defined as 
 \cite{schnakenberg,maes1,esposito}

\begin{eqnarray}
  \dot{S}_{p}=  \frac{1}{2}
  \sum_{nk}(w_{nk}p_{k}-w_{kn}p_{n})
  \ln\frac{w_{nk}p_{k}}{w_{kn}p_{n}}.
\end{eqnarray}
The inequality (26) is similar to the so-called speed limit relation found
for the evolution of Markov thermodynamic systems \cite{shiraishi}.

The term
$\langle|\big(p/q\big)^{\alpha-1} - C_{\alpha}|^{3}\rangle_{p}$
in Eq. (23)
is bounded by (see Eq. (C1) in the Appendix C):

\begin{eqnarray}
  \langle|\big(p/q\big)^{\alpha-1} - C_{\alpha}|^{3} \rangle_{p}
  \le  C_{3\alpha-2} -C_{\alpha}C_{2\alpha-1}.
\end{eqnarray}
Combining Eqs. (23), (24), (26) and (28) we obtain either

\begin{eqnarray}
  \langle |\dot{p}/p|
  |\big(p/q\big)^{\alpha-1} - C_{\alpha}| \rangle_{p}
\le \Big( 2I_{F}(p) A_{p}
  \nonumber  \\
\times \big[C_{3\alpha-2} -C_{\alpha}C_{2\alpha-1}\big] \Big)^{1/3},
\end{eqnarray}         

or

\begin{eqnarray}
  \langle |\dot{p}/p|
  |\big(p/q\big)^{\alpha-1} - C_{\alpha}| \rangle_{p}
\le \Big( I_{F}(p) \sqrt{2\dot{S}_{p}A_{p}}
  \nonumber  \\
\times \big[C_{3\alpha-2} -C_{\alpha}C_{2\alpha-1}\big] \Big)^{1/3}.
\end{eqnarray}         
Analogical inequalities can be obtained for the remaining term in the
last line of Eq. (6), with appropriately defined $\dot{S}_{q}$ and
$A_{q}$ for the distribution $q$. Taking all that into account leads to
the limits on the rate of $\alpha$-coefficient, as

\begin{eqnarray}
 | \frac{dC_{\alpha}}{dt}| \le |\alpha|\Big(2I_{F}(p) A_{p}
 \big(C_{3\alpha-2} -C_{\alpha}C_{2\alpha-1}\big)\Big)^{1/3}
  \nonumber  \\    
+ |\alpha-1| \Big(2I_{F}(q) A_{q}
 \big(C_{3\alpha} -C_{\alpha}C_{2\alpha}\big)\Big)^{1/3},
  \nonumber  \\    
\end{eqnarray}         
which is strictly kinematic bound with the right hand side called from
now on the bound B3, and
\begin{eqnarray}
 | \frac{dC_{\alpha}}{dt}| \le |\alpha|\Big(I_{F}(p)\sqrt{2\dot{S}_{p}A_{p}}
 \big[C_{3\alpha-2} -C_{\alpha}C_{2\alpha-1}\big]\Big)^{1/3}
  \nonumber  \\    
+ |\alpha-1| \Big(I_{F}(q) \sqrt{2\dot{S}_{q}A_{q}}
 \big[C_{3\alpha} -C_{\alpha}C_{2\alpha}\big]\Big)^{1/3},
  \nonumber  \\    
\end{eqnarray}
which represents a mixed kinematic-thermodynamic bound called the bound B4.
These two equations constitute the third major result of this paper. They
mean that for Markov dynamics $dC_{\alpha}/dt$ can be bounded not only by
the global rate of system dynamics ($I_{F}(p), I_{F}(q)$), but also by average
activities ($A_{p}, A_{q}$), and/or the thermodynamic entropy production
rates ($\dot{S}_{p}, \dot{S}_{q}$). This generally suggests that
kinematics characteristics of the stochastic system are at least as important
as the entropic (energetic) characteristics, in agreement with the
notions in Ref. \cite{maes2}. Furthermore, the restriction to Markov dynamics
makes the bounds in Eqs. (31) and (32) less general than the bounds
in Eqs. (12) and (18).

\subsubsection{Rates of Tsallis and Renyi divergences.}

The inequalities in Eqs. (31) and (32) allow us to write the corresponding
kinematic and thermodynamic bounds on the rate of Tsallis and Renyi divergences.
For the $dT_{\alpha}/dt$ we get the following inequality

\begin{widetext}
\begin{eqnarray}
  | \frac{dT_{\alpha}}{dt}| \le \frac{|\alpha|}{|\alpha-1|^{2/3}}
  \Big(I_{F}(p)\sqrt{A_{p}\Psi_{p}}
\big[3T_{3\alpha-2} -2T_{2\alpha-1}[1+(\alpha-1)T_{\alpha}]- T_{\alpha}\big]\Big)^{1/3}
  \nonumber  \\    
+ \Big(I_{F}(q) \sqrt{A_{q}\Psi_{q}}
\big[(3\alpha-1)T_{3\alpha} -(2\alpha-1)T_{2\alpha}[1+(\alpha-1)T_{\alpha}]
 - (\alpha-1)T_{\alpha}\big]\Big)^{1/3},
\end{eqnarray}
%\end{widetext}
and for $dR_{\alpha}/dt$ we have

\begin{eqnarray}
  | \frac{dR_{\alpha}}{dt}| \le \frac{|\alpha|}{|\alpha-1|}
  \Big(I_{F}(p)\sqrt{A_{p}\Psi_{p}}
 \big[ e^{3(\alpha-1)(R_{3\alpha-2}-R_{\alpha})} -e^{2(\alpha-1)(R_{2\alpha-1}-R_{\alpha})}\big] \Big)^{1/3}
  \nonumber  \\    
+ \Big(I_{F}(q) \sqrt{A_{q}\Psi_{q}}
\big[ e^{(3\alpha-1)R_{3\alpha}-3(\alpha-1)R_{\alpha}} - e^{(2\alpha-1)R_{2\alpha}-2(\alpha-1)R_{\alpha}}
\big] \Big)^{1/3},
%  \nonumber  \\    
\end{eqnarray}         
\end{widetext}
where $\Psi_{\gamma}$, with index $\gamma$ either $p$ or $q$, is

\begin{eqnarray}
%\begin{displaymath}
  \Psi_{\gamma} = \left\{
  \begin{array}{cl}
    4A_{\gamma},  &\;  \mbox{K bound}  \\
    2\dot{S}_{\gamma},  &\;  \mbox{KT bound},
  \end{array}
  \right.
%\end{displaymath}
\end{eqnarray}
with K and KT denoting respectively purely kinematic and mixed
kinematic-thermodynamic bounds.

Eqs. (33) and (34) are slightly more complicated than the basal Eqs. (31) and (32),
chiefly by the presence of various combinations of Tsallis and Renyi divergences of
different order. However, in the case when the distribution of $q$ is a steady-state
distribution of $p$, the terms proportional to $I_{F}(q)$ vanish, and Eqs. (33) and
(34) take simpler forms. For instance, for $\alpha=2$, we obtain the
limit on the rate of Pearson divergence as
$|dT_{2}/dt| \le 2 \Big(I_{F}(p)\sqrt{A_{p}\Psi_{p}}
\big[3T_{4} -2T_{3}(1+T_{2})- T_{2}\big]\Big)^{1/3}$.

\subsubsection{ Rate of Kullback-Leibler divergence.}

Now we turn to the rate of Kullback-Leibler divergence, with Eq. (7) as the
starting point. Applying the H\"older inequality in the same way as above, we get

\begin{eqnarray}
  \langle |\dot{p}/p|
  |\ln(p/q) - D_{KL}| \rangle_{p}
  \nonumber  \\
\le \Big( I_{F}(p) \langle|\dot{p}/p|\rangle_{p}
  \langle |\ln(p/q) - D_{KL}|^{3}\rangle_{p}\Big)^{1/3}.
\end{eqnarray}         
and
\begin{eqnarray}
  \langle |\dot{q}/q| |(p/q) - 1| \rangle_{q}
%  \nonumber  \\
\le \Big( I_{F}(q) \langle|\dot{q}/q|\rangle_{q}
  \langle |(p/q) - 1|^{3}\rangle_{q}\Big)^{1/3}.
 \nonumber  \\
\end{eqnarray}         
The terms $\langle |\dot{p}/p|\rangle_{p}$ and $\langle |\dot{q}/q|\rangle_{q}$ 
are restricted by Eqs. (24) and (26) or their analogs. 

The bound on  $\langle|(p/q) - 1|^{3}\rangle_{q}$ is obtained immediately
from Eq. (C1), with the result

\begin{eqnarray}
  \langle|(p/q) - 1|^{3} \rangle_{q}  \le  C_{3} - C_{2}= 2T_{3} - T_{2}.
\end{eqnarray}

Estimating  $\langle |\ln(p/q) - D_{KL}|^{3}\rangle_{p}$
requires more transformations. With the help of Eq. (C5) in the
Appendix C that term can be bounded by various $\alpha$-coefficients as

\begin{eqnarray}
 \langle |\ln(p/q) - D_{KL}|^{3}\rangle_{p}
 \le  e^{-3D_{KL}/2}C_{5/2} - e^{-D_{KL}/2}C_{3/2}
 \nonumber \\
 - e^{D_{KL}/2}C_{1/2} + e^{3D_{KL}/2}C_{-1/2}.
 \nonumber \\
\end{eqnarray}

Combining Eqs. (7), (24), (26) and (36-39), we obtain
the limit on the rate of KL divergence

\begin{eqnarray}
 |\frac{dD_{KL}}{dt}|  \le
 \Big(I_{F}(q) \sqrt{A_{q}\Psi_{q}}
 \big[2T_{3} -T_{2}\big]\Big)^{1/3}
  \nonumber  \\    
+ \Big(I_{F}(p)\sqrt{A_{p}\Psi_{p}}
 \big[ e^{-3D_{KL}/2}C_{5/2} - e^{-D_{KL}/2}C_{3/2}
  \nonumber \\
 - e^{D_{KL}/2}C_{1/2} + e^{3D_{KL}/2}C_{-1/2}\big] \Big)^{1/3},
  \nonumber  \\    
\end{eqnarray}
where the quantities $\Psi_{p}$ and $\Psi_{q}$ are given by Eq. (35).
As can be seen, apart from similar terms as those in Eqs. (33) and (34),
the upper bound contains also various exponents of $D_{KL}$.
More broadly, one can interpret the kinematic-thermodynamic bounds in Eqs.
(33), (34) and (40) that the predictability of the system dynamics
is associated with its levels of dissipation and dynamical agitation.
The smaller these two factors, the better is the prediction of the dynamics.

\subsection{\label{sec:levell3} Bounds of second kind.}

In this section we derive a second, alternative, thermodynamic-kinematic bound
on the rate of statistical divergences.

Consider the first line of Eq. (6). We can substitute for $\dot{p}_{n}$ and
$\dot{q}_{n}$ in this equation their Master equation dynamics given by Eq. (22).
This leads to

\begin{eqnarray}
 | \frac{dC _{\alpha}}{dt}| \le
  |\alpha| \sum_{nk} |w_{nk}p_{k}-w_{kn}p_{n}| |\big(p_{n}/q_{n}\big)^{\alpha-1} - C_{\alpha}|
  \nonumber  \\   
  + |\alpha-1| \sum_{nk} |v_{nk}q_{k}-v_{kn}q_{n}| |\big(p_{n}/q_{n}\big)^{\alpha} - C_{\alpha}|.
  \nonumber  \\
\end{eqnarray}         
The first term on the right proportional to $|\alpha|$ can be limited again
in two different ways (see Eq. (B6) in Appendix B):

\begin{eqnarray}
  \sum_{nk} |w_{nk}p_{k}-w_{kn}p_{n}| |\big(p_{n}/q_{n}\big)^{\alpha-1} - C_{\alpha}|
%  \nonumber  \\
  \le \sqrt{\Psi_{p}/2}
\nonumber  \\
 \times \Big(\sqrt{I_{F}(p)} + 2\sqrt{\langle\overline{w}^{2}\rangle_{p}}\Big)^{1/2}
  \langle \big[(p/q)^{\alpha-1} - C_{\alpha}\big]^{4}\rangle_{p}^{1/4},
  \nonumber  \\
\end{eqnarray}
where  $\langle\overline{w}^{2}\rangle_{p}= \sum_{n} \overline{w}_{n}^{2}p_{n}$
is the second moment of total escape rate.

Moreover, by a direct computation we have

\begin{eqnarray}
  \langle \big[(p/q)^{\alpha-1} - C_{\alpha}\big]^{4}\rangle_{p}
  = C_{4\alpha-3} - 4C_{\alpha}C_{3\alpha-2}
  \nonumber  \\
  +  6C_{\alpha}^{2}C_{2\alpha-1} - 3C_{\alpha}^{4}.  
\end{eqnarray}

Combining Eqs. (41-43), and applying the same reasoning for the second term
in Eq. (41), we obtain two upper bounds on the rate of $\alpha$-coefficient
depending on the value for $\Psi$:

\begin{widetext}
\begin{eqnarray}
 | \frac{dC _{\alpha}}{dt}| \le
 |\alpha|\sqrt{\Psi_{p}/2}
\Big(\sqrt{I_{F}(p)} + 2\sqrt{\langle\overline{w}^{2}\rangle_{p}}\Big)^{1/2}
%\nonumber  \\
 \Big( C_{4\alpha-3} - 4C_{\alpha}C_{3\alpha-2}
  +  6C_{\alpha}^{2}C_{2\alpha-1} - 3C_{\alpha}^{4}\Big)^{1/4}
   \nonumber  \\
+ |\alpha-1|\sqrt{\Psi_{q}/2}
\Big(\sqrt{I_{F}(q)} + 2\sqrt{\langle\overline{v}^{2}\rangle_{q}}\Big)^{1/2}
%\nonumber  \\
 \Big( C_{4\alpha} - 4C_{\alpha}C_{3\alpha}
  +  6C_{\alpha}^{2}C_{2\alpha} - 3C_{\alpha}^{4}\Big)^{1/4}. 
\end{eqnarray}
\end{widetext}
This equation is an alternative to Eqs. (31) and (32), though a little more
complicated, and combines a purely kinematic bound called B5 (for $\Psi= 4A$)
with a mixed kinematic-thermodynamic bound called B6 (for $\Psi= 2\dot{S}$).
Note that in the steady state for both probability distributions $p$ and $q$, all
the terms $dC_{\alpha}/dt$, $\dot{S}_{p}$, $\dot{S}_{q}$, $I_{F}(p)$,
and $I_{F}(q)$ are zero, but $\langle\overline{w}^{2}\rangle_{p}$ and
$\langle\overline{v}^{2}\rangle_{p}$ are nonzero.

The corresponding bounds on the rates of Tsallis and Renyi divergences can
be obtained straightforwardly from Eq. (44), using transformation in Eqs. (2)
and (3). The resulting inequalities are similar to Eq. (44), although more
elaborate due to more complicated combination of $\alpha$-coefficients.
Below, instead, we provide an explicit bound on the rate of KL divergence,
which takes the form

%\begin{widetext}
\begin{eqnarray}
 | \frac{dD _{KL}}{dt}| \le
 \sqrt{\Psi_{p}/2}
\Big(\sqrt{I_{F}(p)} + 2\sqrt{\langle\overline{w}^{2}\rangle_{p}}\Big)^{1/2}
\nonumber  \\
\times \big(\langle \big[\ln(p/q)-D_{KL}\big]^{4}\rangle_{p}\big)^{1/4}
   \nonumber  \\
+ \sqrt{\Psi_{q}/2}
\Big(\sqrt{I_{F}(q)} + 2\sqrt{\langle\overline{v}^{2}\rangle_{q}}\Big)^{1/2}
\nonumber  \\
\times \Big(3T_{4} - 8T_{3}  +  6T_{2}\Big)^{1/4},
\end{eqnarray}
%\end{widetext}
where  $\langle \big[\ln(p/q)-D_{KL}\big]^{4}\rangle_{p}$  can
be bounded as (see Eq. (C6) in Appendix C)

\begin{eqnarray}
 \langle [\ln(p/q) - D_{KL}]^{4}\rangle_{p}
 \le  e^{-2D_{KL}}C_{3} - 4e^{-D_{KL}}C_{2} + 6
 \nonumber \\
 - 4e^{D_{KL}} + e^{2D_{KL}}C_{-1}.
 \nonumber \\
\end{eqnarray}

Together Eqs. (44-46) constitute the forth major result of this work.
They imply that the rate of information gain about the system dynamics
is restricted by both thermodynamic and kinematic characteristics, both of
the true system (probabilities $q$) and its estimator (probabilities $p$).

\section{\label{sec:level3} Comparison of the bounds: One-step driven process.}

In Table 1 all the derived bounds on the rates of statistical divergences are summarized.

Next, we check the quality of the six upper bounds on $dC_{\alpha}/dt$,
denoted by B1-B6 and represented by Eqs. (12), (18), (31), (32), and (44),
respectively. We choose a specific example of a stochastic dynamical system
known as the one-step Markov jump process (known also as birth-and-death process)
with $N+1$ states \cite{vankampen}. We consider two versions of this system: one
driven by periodic stimulation, and another relaxing to its steady state.
For the driven case, the probability $p_{n}$ of being in state $n$ is described
by the following master equation

\begin{eqnarray*}
  \dot{p}_{n}= w_{n,n-1}p_{n-1} + w_{n,n+1}p_{n+1}  
  \nonumber  \\
  -(w_{n-1,n} + w_{n+1,n})p_{n},
\end{eqnarray*}
for $n=1,...,N-1$, and for the boundary probabilities we have
$\dot{p}_{0}= w_{0,1}p_{1} - w_{1,0}p_{0}$,
and $\dot{p}_{N}= w_{N,N-1}p_{N-1} - w_{N-1,N}p_{N}$, with the transition rates
$w_{n-1,n}= a_{0}n$, $w_{n+1,n}= b(t)(N-n)$, where the time-dependent
oscillating rate $b(t)= b_{0}\big(1 + g[\cos(\omega t)+1]\big)$.
The parameters $a_{0}$ and $b_{0}$ are respectively the amplitudes of the
downhill and uphill transitions, and oscillations of $b(t)$ are controlled by
amplitude $g$ and frequency $\omega$.

For the relaxing case, we have the same structure of the master equation as above,
but we denote the corresponding probabilities as $q_{n}$, with the time-independent
transitions rates $v_{n-1,n}= w_{n-1,n}$ and $v_{n+1,n}= b_{0}(N-n)$. In both cases,
the same initial condition on the probabilities was used, i.e.,
$p_{i}(0)= q_{i}(0)= \frac{1}{N+1}$ for all $i=0,1,...,N$, which means that
initially all the states are equally likely.

For this system we compute numerically the $\alpha$-coefficient $C_{\alpha}(p||q)$
and its time derivative, as well as all B1-B6 bounds on $dC_{\alpha}/dt$.
Overall, the best estimate for $|dC_{\alpha}/dt|$ is provided by the bound B1,
and the discrepancy between the two is very small as time progresses (Figs. 1 and 2).
The bounds B2, B4 and B6 compete for the second place, but their ranking
changes dynamically. Their mutual relationship depends also on the order $\alpha$
(compare Figs. 1 and 2).
The kinematic bounds B3 and B5 are rather weak, especially B5 which does
not fit into the scale of Figs. 1 and 2. This suggests that purely kinematic
bounds for Markov processes obeying master equation do not capture well
the rates of statistical divergences.

The superiority of the general kinematic bound B1 (Eq. 12) follows from two facts.
The first is that its derivation involves a minimal number of mathematical
transformations, i.e., less consequitive inequalities on the way is required, and
thus less inaccuracies is introduced. The second reason is more subtle and it
concerns the number of constraints on the physical variables, appearing in
the bounds, which have to be satisfied to make the bounds the good estimates
of $|dC_{\alpha}/dt|$. The larger the number of constraints the less likely the
bound will be reached. To be more specific, let us compare the bound B1
with the bounds B3 and B4. The bound B1 is derived from the H\"older
inequality (Eq. 4) for $m=2$, i.e., only two variables are involved.
Consequently, this inequality becomes equality when only one constraint
relating the two variables is satisfied
($|X_{1}|^{1/\lambda_{1}} \sim |X_{2}|^{1/\lambda_{2}}$). The equality in the
H\"older inequality corresponds to the saturation of the bound B1.
On the other hand, the bounds B3 and B4 (Eqs. 31 and 32) are derived
from Eq. (4) with $m=3$ variables. To achieve equality in Eq. (4) in this
case requires two constraints on these three variables
($|X_{1}|^{1/\lambda_{1}} \sim |X_{2}|^{1/\lambda_{2}} \sim |X_{3}|^{1/\lambda_{3}}$),
which is much more restrictive on the dynamics of these variables than in the
former case. As a result the bounds B3 and B4 are more difficult to reach,
and their values deviate significantly from the actual value of 
$|dC_{\alpha}/dt|$.

\begin{widetext}
\begin{table}
\begin{center}
\caption{Summary of the inequalities for the rates of divergences.}
\begin{tabular}{|l c c |}
\hline

Bound    &  Divergence  & Equation        \\  
         &  type        &                \\

\hline

{\bf B1}  &           &       \\
 & Chernoff  &
$|\dot{C}_{\alpha}| \le
  |\alpha|\sqrt{I_{F}(p)}\sqrt{C_{2\alpha-1}-C_{\alpha}^{2}}
  +  |\alpha-1|\sqrt{I_{F}(q)}\sqrt{C_{2\alpha}-C_{\alpha}^{2}} $  \\
       
 & Tsallis  &
 $|\dot{T}_{\alpha}| \le
 |\alpha| \sqrt{I_{F}(p)}\Big[\frac{2}{(\alpha-1)}(T_{2\alpha-1}-T_{\alpha})-T_{\alpha}^{2}\Big]^{1/2}
  +  \sqrt{I_{F}(q)}\big[(2\alpha-1)T_{2\alpha}-2(\alpha-1)T_{\alpha}
    - (\alpha-1)^{2}T_{\alpha}^{2}\big]^{1/2} $    \\

    &  Renyi  &     
  $|\dot{R}_{\alpha}| \le \frac{|\alpha|}{|\alpha-1|}
    \sqrt{I_{F}(p)}\big[e^{2(\alpha-1)(R_{2\alpha-1}-R_{\alpha})} - 1\big]^{1/2}
  +  \sqrt{I_{F}(q)}\big[e^{[(2\alpha-1)R_{2\alpha}-2(\alpha-1)R_{\alpha}]} - 1\big]^{1/2}$   \\

    &   Kullback-Leibler  &
 $|\dot{D}_{KL}| \le
  \sqrt{I_{F}(p)}\sqrt{T_{2} - D_{KL}^{2}}
  +  \sqrt{I_{F}(q)}\sqrt{T_{2}} $ \\

    &           &        \\
 {\bf B2}  &           &       \\
  
& Chernoff  &
$ |\dot{C}_{\alpha}| \le   \sqrt{I_{F}(q)} \sqrt{C_{2\alpha}- C_{\alpha}^{2}}   
+ |\alpha|\sqrt{F(p||q)} \sqrt{C_{2\alpha-1}}$  \\  

& Tsallis  &
 $|\dot{T}_{\alpha}| \le
  \sqrt{I_{F}(q)} \big[\frac{(2\alpha-1)T_{2\alpha}-2(\alpha-1)T_{\alpha}}
    {(\alpha-1)^{2}} - T_{\alpha}^{2}\big]^{1/2}
+ \frac{|\alpha|}{|\alpha-1|}\sqrt{F(p||q)} \big[2(\alpha-1)T_{2\alpha-1}+1\big]^{1/2}$ \\

&  Renyi  &     
 $ |\dot{R}_{\alpha}| \le \frac{\sqrt{I_{F}(q)}}{|\alpha-1|}
  \sqrt{e^{(2\alpha-1)R_{2\alpha}-2(\alpha-1)R_{\alpha}} - 1}
+ \frac{|\alpha|}{|\alpha-1|}\sqrt{F(p||q)} e^{(\alpha-1)(R_{2\alpha-1}-R_{\alpha})} $ \\

&   Kullback-Leibler  &
 $|\dot{D}_{KL}| \le
  \sqrt{I_{F}(p)} \sqrt{T_{2} - D_{KL}^{2}}   
  + \sqrt{F(p||q)}$   \\

        &           &        \\
{\bf B3, B4}  &           &       \\
& Chernoff  &
  $ |\dot{C}_{\alpha}| \le |\alpha|\Big(I_{F}(p)\sqrt{A_{p}\Psi_{p}}
 \big[C_{3\alpha-2} -C_{\alpha}C_{2\alpha-1}\big]\Big)^{1/3}
+ |\alpha-1| \Big(I_{F}(q) \sqrt{A_{q}\Psi_{q}}
 \big[C_{3\alpha} -C_{\alpha}C_{2\alpha}\big]\Big)^{1/3} $   \\

& Tsallis  &
  $|\dot{T}_{\alpha}| \le \frac{|\alpha|}{|\alpha-1|^{2/3}}
  \big(I_{F}(p)\sqrt{A_{p}\Psi_{p}}
[3T_{3\alpha-2} -2T_{2\alpha-1}[1+(\alpha-1)T_{\alpha}]- T_{\alpha}]\big)^{1/3}$ \\
 &   &
$ + \big(I_{F}(q) \sqrt{A_{q}\Psi_{q}}
[(3\alpha-1)T_{3\alpha} -(2\alpha-1)T_{2\alpha}[1+(\alpha-1)T_{\alpha}]
 - (\alpha-1)T_{\alpha}]\big)^{1/3} $  \\

&  Renyi  &     
  $| \dot{R}_{\alpha}| \le \frac{|\alpha|}{|\alpha-1|}
  \big(I_{F}(p)\sqrt{A_{p}\Psi_{p}}
 [ e^{3(\alpha-1)(R_{3\alpha-2}-R_{\alpha})} -e^{2(\alpha-1)(R_{2\alpha-1}-R_{\alpha})}] \big)^{1/3}$ \\
 &   &
$ + \big(I_{F}(q) \sqrt{A_{q}\Psi_{q}}
[ e^{(3\alpha-1)R_{3\alpha}-3(\alpha-1)R_{\alpha}} - e^{(2\alpha-1)R_{2\alpha}-2(\alpha-1)R_{\alpha}}] \big)^{1/3}$   \\

&   Kullback-Leibler  &
$ |\dot{D}_{KL}|  \le
 \big(I_{F}(q) \sqrt{A_{q}\Psi_{q}}[2T_{3} -T_{2}]\big)^{1/3}    
+ \big(I_{F}(p)\sqrt{A_{p}\Psi_{p}}\big)^{1/3}$   \\  
 &  &
$\times\big[ e^{-3D_{KL}/2}C_{5/2} - e^{-D_{KL}/2}C_{3/2}
 - e^{D_{KL}/2}C_{1/2} + e^{3D_{KL}/2}C_{-1/2}\big]^{1/3}$ \\

        &           &        \\
{\bf B5, B6}  &           &       \\
& Chernoff  &
$ |\dot{C} _{\alpha}| \le  |\alpha|\sqrt{\Psi_{p}/2}
\big(\sqrt{I_{F}(p)} + 2\sqrt{\langle\overline{w}^{2}\rangle_{p}}\big)^{1/2}
 \big( C_{4\alpha-3} - 4C_{\alpha}C_{3\alpha-2}
  +  6C_{\alpha}^{2}C_{2\alpha-1} - 3C_{\alpha}^{4}\big)^{1/4} $ \\
&  &   
$  + |\alpha-1|\sqrt{\Psi_{q}/2}
\big(\sqrt{I_{F}(q)} + 2\sqrt{\langle\overline{v}^{2}\rangle_{q}}\big)^{1/2}
 \big( C_{4\alpha} - 4C_{\alpha}C_{3\alpha}
  +  6C_{\alpha}^{2}C_{2\alpha} - 3C_{\alpha}^{4}\big)^{1/4} $  \\

& Tsallis  &  $-$   \\

&  Renyi  &   $-$   \\

&   Kullback-Leibler  &
$ |\dot{D} _{KL}| \le \sqrt{\Psi_{p}/2}
\big(\sqrt{I_{F}(p)} + 2\sqrt{\langle\overline{w}^{2}\rangle_{p}}\big)^{1/2}
\big(e^{-2D_{KL}}C_{3} - 4e^{-D_{KL}}C_{2} + 6 - 4e^{D_{KL}} + e^{2D_{KL}}C_{-1}\big)^{1/4} $  \\
&  &
$ + \sqrt{\Psi_{q}/2}
\big(\sqrt{I_{F}(q)} + 2\sqrt{\langle\overline{v}^{2}\rangle_{q}}\big)^{1/2}
 \big(3T_{4} - 8T_{3}  +  6T_{2}\big)^{1/4}$   \\

\hline
\end{tabular}
\end{center}

$\Psi_{\gamma}$ is equal either $4A_{\gamma}$ or $2\dot{S}_{\gamma}$.  

\end{table}
\end{widetext}

It is also interesting to consider why the bound B3 is much less
accurate than the bound B4, given that they both follow from the
similar derivation scheme with $m=3$ in the H\"older inequality.
Since the bounds B3 and B4 differ only by the factor $\Psi_{\gamma}$
in Eq. (35), the fact that B3 gives much larger values than B4 means
that entropy production $\dot{S}_{\gamma}$ is smaller than activity $A_{\gamma}$. 
This seems reasonable because $\dot{S}_{\gamma}$ can be close to 0 for
systems close to equilibrium, while activity $A_{\gamma}$ is always
strictly positive and can be large regardless of the distance from
equilibrium \cite{maes2}.
Thus, in this particular case, having more information about the system
(both activity and entropy production than activity only), is more
adventageous and produces a tighter bound. This is similar to the
case of improved thermodynamic uncertainty relation \cite{vu,vo}.
However, this is not a general rule, as the case of B1 bound vs. B3,B4
bounds shows. For the former bound we have less specific information
about the system, and yet that bound is shown to perform the best.

\begin{figure}[h!]
    \centering
    \includegraphics[scale=0.49]{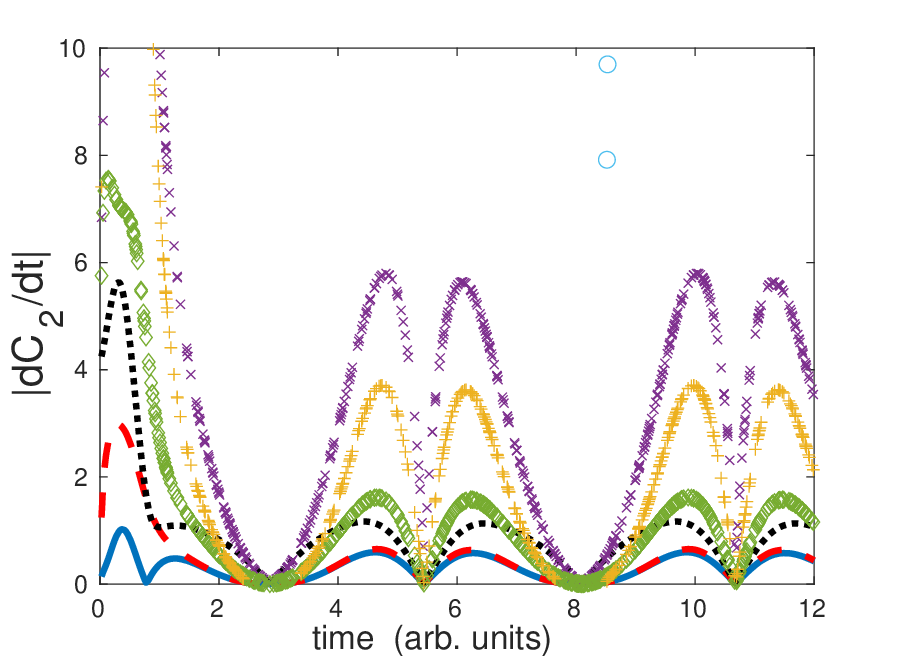}
    \caption{Rate of $\alpha$-coefficient in comparison to its various
      upper bounds as functions of time for $\alpha=2$, corresponding
      to Pearson divergence. Solid line (blue) corresponds to exact value of
      $|dC_{2}/dt|$ and was computed from Eq. (6).
      Upper bounds on $|dC_{2}/dt|$, i.e., B1-B6 are shown as:
      dashed (red) line for B1; dotted (black) line for B2; crosses (purple)
      for B3; diamonds (green) for B4; circles (light blue) for B5; and
      pluses (yellow) for B6.
      Note that the best estimates for $|dC_{2}/dt|$ are given by the
      kinematic bounds B1 (dashed line) and B2 (dotted line), but the
      former is closer to the actual value of $|dC_{2}/dt|$. The bound
      B5 provides a poor estimate and is mostly out of scale.
      Parameters used: $a=a_{0}=3.0$, $b_{0}=1.0$, $g=0.7$, $\omega=1.2$,
      $N= 9$.}
%    \label{fig:enter-label}
\end{figure}

\begin{figure}[h!]
    \centering
    \includegraphics[scale=0.49]{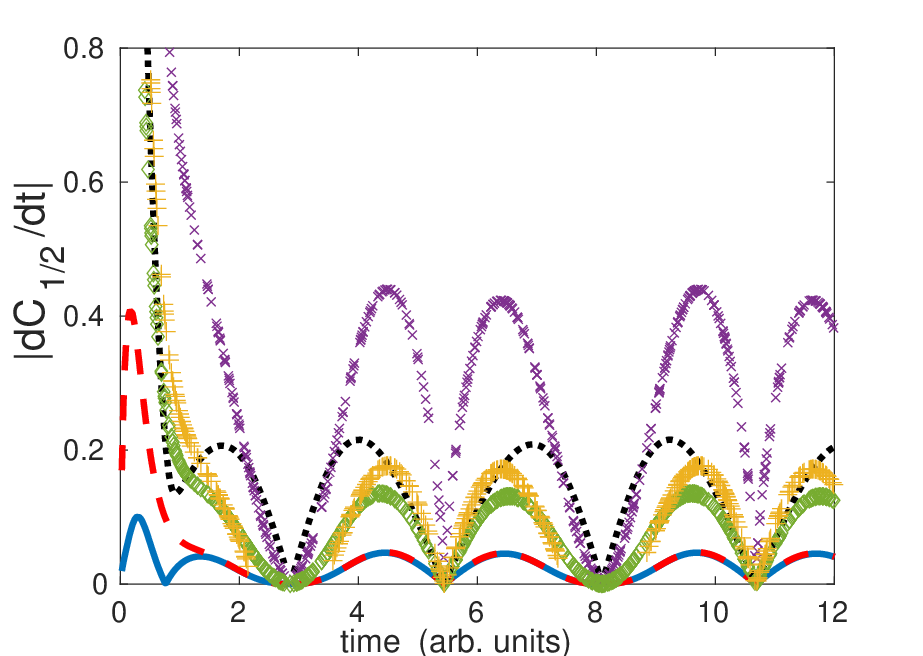}
    \caption{The same as in Fig. 1 but for $\alpha=1/2$, corresponding
      to Hellinger distance. Again the bound B1 gives the best estimate,
      but the accuracy for some other bounds is different than in Fig. 1.
      Notably, the bounds B4 (diamonds) and B6 (pluses) provide
      often better estimates than the bound B2 (dotted line).
      The same labels used and parameters as in Fig. 1.}
%    \label{fig:enter-label}
\end{figure}

\section{\label{sec:level3} Applications of inequalities for the rates of divergences}

The above inequalities for the upper bounds of various divergences can be used
in different circumstances encountered in physics and interdisciplinary research.
Having the bound on the modul of divergence rate $\dot{D}_{\alpha}$ ($= dD_{\alpha}/dt$),
where $D_{\alpha}$ is either Tsallis $T_{\alpha}$ or Renyi divergence $R_{\alpha}$,
allows us to find the discrepancy between $D_{\alpha}$ at different time moments.
In particular, because of the general relationship

\begin{eqnarray}
  D_{\alpha}(T) - D_{\alpha}(0) = \int_{0}^{T} \dot{D}_{\alpha} dt
\nonumber  \\ 
\le  \sqrt{T}\sqrt{\int_{0}^{T} |\dot{D}_{\alpha}|^{2} dt},
\end{eqnarray}         
and because of the upper bounds on $|\dot{D}_{\alpha}|$, we can estimate the
maximal difference between the values of divergences at initial and
some later arbitrary time $T$.

It is worth to stress that the basic inequalities on the rates of statistical
divergences (Eqs. 12-15) have a very similar structure to the inequalities for
the average rates of stochastic observables \cite{ito,nicholson}. Both of them
follow from the Cauchy-Schwartz inequality and contain temporal Fisher information.
In our case, the role of the observable is played by the statistical divergence,
which can be interpreted as a generalized information gain.

\subsection{\label{sec:levell3} Minimal speed and entropy production
  in terms of the rates of statistical divergences.}

In recent years, different speed limits on stochastic thermodynamics in different
systems have been found \cite{vu,shiraishi,margolus,shanahan,hamazaki,garcia,deffner}.
Similarly, there has been an interest in determining minimal entropy production
during stochastic evolution \cite{vu,barato2015,li,vu2020,skinner,salazar,dechant}.
Here, we provide alternative lower limits on the speed of dynamical systems and
their entropy production, using statistical divergences.

In a particular case when divergence $D_{\alpha}(p|p_{\infty})$ is between
time-dependent system's probability distribution $p$ and its steady state
distribution $p_{\infty}$, the divergence $D_{\alpha}(p|p_{\infty})$ can be
interpreted as a generalized information gain in relation to its steady state.
Consequently, the rate of divergence $\dot{D}_{\alpha}(p|p_{\infty})$ can be
thought as the speed of information gaining away from the steady state.

The speed of global system dynamics can be defined as a square root of
temporal Fisher information, i.e., $\sqrt{I_{F}(p)}$. Thus, Eqs. (13) and
(14) provide lower bounds on the speed of system evolution through either
Tsallis or Renyi divergences, or $\alpha$- coefficient as

\begin{eqnarray}
\sqrt{I_{F}(p)}  \ge
 \frac{|\dot{C}_{\alpha}(p||p_{\infty})|} 
{ |\alpha| 
  \sqrt{C_{2\alpha-1}-C_{\alpha}^{2}} },
\end{eqnarray}
\begin{eqnarray}
\sqrt{I_{F}(p)}  \ge
 \frac{|\dot{T}_{\alpha}(p||p_{\infty})|} 
{ |\alpha| 
  \sqrt{\frac{2}{(\alpha-1)}(T_{2\alpha-1}-T_{\alpha})-T_{\alpha}^{2}} },
\end{eqnarray}
and 
\begin{eqnarray}
\sqrt{I_{F}(p)}  \ge
 \frac{|\alpha-1||\dot{R}_{\alpha}(p||p_{\infty})|} 
{ |\alpha| \sqrt{e^{2(\alpha-1)(R_{2\alpha-1}-R_{\alpha})} - 1} }.
\end{eqnarray}
These inequalities imply that the minimal speed of system's stochastic
dynamics is set by the rate of generalized information gain in this system.

Similarly, we can provide lower bounds on the entropy production rate in
stochastic Markov systems by inverting Eqs. (32-34). As before, by considering
statistical divergences between time dependent distribution $p$ and its steady
state form $p_{\infty}$, we obtain the following inequalities
for $\dot{S}_{p}$:

\begin{eqnarray}
\dot{S}_{p}  \ge
 \frac{\dot{C}_{\alpha}(p||p_{\infty})^{6}} 
{ 2\alpha^{6}A_{p} I_{F}(p)^{2}[C_{3\alpha-2}-C_{\alpha}C_{2\alpha-1}]^{2} },
\end{eqnarray}
and via the rates of Tsallis and Renyi divergences

\begin{eqnarray}
\dot{S}_{p}  \ge
 \frac{(\alpha-1)^{4}\dot{T}_{\alpha}(p||p_{\infty})^{6}}
      { 2\alpha^{6}A_{p} I_{F}(p)^{2}}
\nonumber  \\ 
\times [3T_{3\alpha-2}-2T_{2\alpha-1}(1+(\alpha-1)T_{\alpha})- T_{\alpha}]^{-2},
\end{eqnarray}
and
\begin{eqnarray}
\dot{S}_{p}  \ge
 \frac{[(\alpha-1)\dot{R}_{\alpha}(p||p_{\infty})]^{6}}
      { 2\alpha^{6}A_{p} I_{F}(p)^{2} }
\nonumber  \\ 
\times [e^{3(\alpha-1)(R_{3\alpha-2}-R_{\alpha})} - e^{2(\alpha-1)(R_{2\alpha-1}-R_{\alpha})} ]^{-2}.
\end{eqnarray}
Eqs. (51-53) determine minimal dissipation for stochastic thermodynamic
systems in terms of the rates of generalized information gains, system's
average activity, and its speed. As such, they are alternatives to the minimal
limits on entropy production derived in other ways, and involving other quantities
\cite{vu,barato2015,li,vu2020,skinner,salazar,dechant}. Interestingly, the minimal
dissipation in the system is inversely proportional to the product of
the system's activity and the forth power of its speed. Thus, paradoxically,
it is possible, in principle, to increase dissipation by decreasing activity
$A_{p}$ and global speed $\sqrt{I_{F}}$, as long as the rate of information
gain is fixed.

Other, more specific, applications of the rates of divergences are provided below.

\subsection{\label{sec:levell3} Applications in physics and biophysics.}

\subsubsection{Limits on nonequilibrium rates of free energy and work in thermodynamics.}

Let us consider a physical system in contact with an environment (heat bath)
at temperature $T$. Our goal is to find the bounds on the rates of available free
energy and work associated with this system, which can be in thermal equilibrium
or in nonequilibrium with the environment. In the first case, we describe the system
by the probability distribution $p_{eq,n}(t)$, that it is in state $n$ at time $t$,
while in the second, nonequilibrium case, we describe our system analogously by the
probability distribution $p_{n}(t)$. The nonequilibrium version of the second law of
thermodynamics for our system is \cite{esposito2011}

\begin{eqnarray}
 \dot{W} - \dot{F}= k_{B}T \dot{S},
\end{eqnarray}  
where $\dot{W}$ is the rate of work performed on the system, $\dot{F}$ is the rate
of nonequilibrium free energy, $k_{B}$ is the Boltzmann constant, and
$k_{B}T \dot{S}$ is the physical entropy production rate in energy units.
Because of the presence of dissipation in the system, which mathematically means
that $\dot{S} \ge 0$, we obtain the second law as  $\dot{W} \ge \dot{F}$,
or equivalently $\Delta W \ge \Delta F$. These inequalities indicate that
the maximal useful work that can be extracted ($-\Delta W$) is at most
equal to the corresponding decrease in nonequilibrium free energy
($-\Delta F$).

The time-dependent nonequilibrium free energy $F(t)$ is related to the time-dependent
equilibrium free energy $F_{eq}(t)$ by \cite{esposito2011,still}

\begin{eqnarray}
F(t) - F_{eq}(t)= k_{B}T D_{KL}(p||p_{eq}),
\end{eqnarray}  
which means that nonequilibrium free energy is always greater than the equilibrium one
by the amount of information needed to specify the nonequilibrium state (quantified
by the KL divergence between the distributions $p_{n}$ and $p_{eq,n}$). The differences
of the rates of these free energies,
$\dot{F} - \dot{F}_{eq}= k_{B}T \dot{D}_{KL}(p||p_{eq})$, are thus restricted by
the bounds on KL divergence, given by Eqs. (15), (17), or (40) and (45).
For example, using Eq. (17), we obtain bounds on the nonequilibrium
free energy rate as

\begin{eqnarray}
|\dot{F} - \dot{F}_{eq}| \le
k_{B}T \Big( \sqrt{I_{F}(p)}
\sqrt{T_{2}-D_{KL}^{2}}
\nonumber  \\ 
+ \sqrt{I_{F}(p_{eq})} \sqrt{T_{2}} \Big).
\end{eqnarray}
This means that the speed with which free energy changes is limited by the speeds
of global dynamics of nonequilibrium and equilibrium versions of the system
(i.e., $I_{F}(p)$ and $I_{F}(p_{eq})$), as well as by the Pearson and
KL divergences between nonequilibrium and equilibrium distributions
(i.e., $T_{2}(p||p_{eq})$ and $D_{KL}(p||p_{eq})$).

Using Eqs. (54) and (56), we can also write the bounds on the
rate of work performed on the system. We obtain

\begin{eqnarray}
 - k_{B}T \Big( \sqrt{I_{F}(p)}\sqrt{T_{2}-D_{KL}^{2}}
  + \sqrt{I_{F}(p_{eq})} \sqrt{T_{2}} \Big) 
\nonumber  \\ 
\le  \dot{W} - \dot{F}_{eq} - k_{B}T \dot{S} 
\nonumber  \\ 
  \le   k_{B}T \Big( \sqrt{I_{F}(p)}\sqrt{T_{2}-D_{KL}^{2}}
  + \sqrt{I_{F}(p_{eq})} \sqrt{T_{2}} \Big). 
\end{eqnarray}
These inequalities allow us to find lower and upper bounds on the rates of
extracted work from ($-\dot{W}$) or done on ($\dot{W}$) the thermodynamic system.
In a particular case when the equilibrium probability distribution is
time independent, i.e. $\dot{p}_{eq,n}= 0$, we obtain a simpler formula
for the maximal extracted work rate:
$- \dot{W} \le   k_{B}T \big(-\dot{S} + \sqrt{I_{F}(p)}
\sqrt{T_{2}-D_{KL}^{2}} \big)$.
That work rate is restricted not only
by the entropy production rate but also by the global speed of
the nonequilibrium state and the difference in Pearson and KL
divergences.

\subsubsection{Overdamped particle in time dependent potential vs. ``target''
  potential.}

This example concerns kinematic bounds on the rates of divergences given by
Eqs. (12-14). Consider a Brownian massless particle moving in 1D with
trajectory $x(t)$ in a stochastic environment with a damping force $-k dx/dt$
($k$ is some positive constant). We study the motion of this particle in
two different external time dependent potentials, either $V_{1}(s_{1}(t))$
or $V_{2}(s_{2}(t))$, which are influenced by two time dependent arbitrary signals
$s_{1}(t)$ and $s_{2}(t)$. We call the potential $V_{2}$ the target or ``desired''
potential, and $V_{1}$ the actual potential. Our goal is to study how fast the
actual trajectory of the particle, corresponding to $V_{1}$ potential, diverges
from the target trajectory corresponding to $V_{2}$ potential.  
For analytical tractability, we assume harmonic
potentials in both cases, i.e., $V_{i}= \frac{1}{2}k\gamma[x-s_{i}(t)]^{2}$,
where $\gamma$ is the inverse of (relaxation) time constant of the system. 
In this case, the signals $s_{i}(t)$ are the centers of the two potentials.

The equation of motion in both potentials is

\begin{eqnarray}
  \dot{x}= -\gamma[x-s_{i}(t)] + \sqrt{2\gamma\sigma^{2}}\eta(t),
\end{eqnarray}  
where $i=1,2$ and corresponds to the case with potential either $V_{1}$
or $V_{2}$, $\sigma$ is the standard deviation of the noise in the system,
and $\eta(t)$ is the delta-correlated Gaussian random variable related to the
noise such that
$\langle \eta(t)\rangle= 0$, and $\langle \eta(t)\eta(t') \rangle= \delta(t-t')$.

The trajectory $x(t)$ of the particle tries to follow the instantaneous value
of the signal $s_{i}(t)$, but it is distracted by the noise, and thus the
particle position is a stochastic variable. This particle dynamics can be
described equivalently by the dynamics of probability density of particle
position in terms of the Fokker-Planck equation as

\begin{eqnarray}
\frac{\partial P_{i}(x,t)}{\partial t}= - \frac{\partial J_{i}(x,t)}{\partial x},
\end{eqnarray} \\
with
\begin{eqnarray}
J_{i}(x,t)= -\gamma[x-s_{i}(t)]P_{i}(x,t)
- \gamma\sigma^{2}\frac{\partial P_{i}(x,t)}{\partial x},
\end{eqnarray}  
where $P_{i}(x,t)$ is the probability density of particle position in the
potential related to signal $s_{i}(t)$, and $J_{i}(x,t)$ is the
probability flux ($i=1,2$). Equation (59) can be exactly solved yielding 
\cite{vankampen}

\begin{eqnarray}
  P_{i}(x,t)= \frac{
    \exp\Big(- \frac{[x-e^{-\gamma t}(x_{0}+\gamma g_{i}(t))]^{2}}
    {2\sigma^{2}(1-e^{-2\gamma t})} \Big) }
  {\sqrt{2\pi\sigma^{2}(1-e^{-2\gamma t})}},
\end{eqnarray} \\  
where $g_{i}(t)= \int_{0}^{t} \; dt' e^{\gamma t'} s_{i}(t')$.
The average value of particle position $x(t)$ in both potentials is
$\langle x(t)\rangle= [x_{0}+\gamma g_{i}(t)]e^{-\gamma t}$, and the
variance is
$\langle[x(t)-\langle x(t)\rangle]^{2}\rangle = \sigma^{2}(1-e^{-2\gamma t})$.

Next, we calculate how the two probability densities $P_{1}(x,t)$
and $P_{2}(x,t)$ diverge as time progresses. For this we find continuous
version of Chernoff $\alpha$-coefficient given by Eq. (1), with
$p(x,t)= P_{1}(x,t)$ and $q(x,t)= P_{2}(x,t)$, i.e.,
$C_{\alpha}(p||q)= \int dx\; p(x,t) \Big[\frac{p(x,t)}{q(x,t)}\Big]^{\alpha-1}$.
We obtain

\begin{eqnarray}
  C_{\alpha}(p||q)= 
  \exp\Big( \frac{\alpha(\alpha-1)\gamma^{2}e^{-2\gamma t}}
  {2\sigma^{2}(1-e^{-2\gamma t})} 
\nonumber  \\ 
\times \Big( \int_{0}^{t} dt' \; e^{\gamma t'} [s_{1}(t')-s_{2}(t')] \Big)^{2}
  \Big),
\end{eqnarray} \\  
from which we can determine the Tsallis and Renyi divergences for the
Brownian particle. For example, the Renyi $R_{\alpha}(p||q)$ divergence
takes the form

\begin{eqnarray}
 R_{\alpha}(p||q)= 
 \frac{\alpha\gamma^{2}e^{-2\gamma t}
 \Big( \int_{0}^{t} dt' \; e^{\gamma t'} [s_{1}(t')-s_{2}(t')] \Big)^{2} }
{2\sigma^{2}(1-e^{-2\gamma t})}, 
\end{eqnarray} \\  
which means that $R_{\alpha}(p||q)$ measures the differences
between the external signals $s_{1}(t)$ and $s_{2}(t)$ accumulated
over time, appropriately weighted.

The rate, at which the distributions $p(x,t)$ and $q(x,t)$ diverge
is given by the derivative $dC_{\alpha}/dt$ and reads

\begin{eqnarray}
 \frac{dC_{\alpha}}{dt}= 
  \frac{ \alpha(\alpha-1)\gamma^{2}e^{-\gamma t}}
       {\sigma^{2}(1-e^{-2\gamma t})}
       \Big( -\frac{\gamma e^{-\gamma t}}{(1-e^{-2\gamma t})}
  [g_{2}(t)-g_{1}(t)]^{2}        
 \nonumber  \\
+  [s_{2}(t)-s_{1}(t)][g_{2}(t)-g_{1}(t)] 
  \Big).
 \nonumber  \\
\end{eqnarray} \\  
Because $dC_{\alpha}/dt$ is a quadratic function of $[g_{2}(t)-g_{1}(t)]$,
it has a maximum proportional to the square of the difference between
the signals $s_{1}(t)$ and $s_{2}(t)$ (for $\alpha > 1$). More formally,

\begin{eqnarray}
 \frac{dC_{\alpha}}{dt}  \le
  \frac{ \alpha(\alpha-1)\gamma C_{\alpha}}
       {4\sigma^{2}} [s_{2}(t)-s_{1}(t)]^{2},
\end{eqnarray} \\
which leads to the maximal rate of Renyi divergence
between the actual particle distribution and its target
distribution

\begin{eqnarray}
 \frac{dR_{\alpha}}{dt}  \le
  \frac{ \alpha\gamma}
       {4\sigma^{2}} [s_{2}(t)-s_{1}(t)]^{2}.
\end{eqnarray} \\

To assess the bounds on the absolute value $|dC_{\alpha}/dt|$
and $|dR_{\alpha}/dt|$, we need
to find the expressions for $C_{2\alpha}-C_{\alpha}^{2}$ and
$C_{2\alpha-1}-C_{\alpha}^{2}$, as well as for temporal Fisher informations.
Using Eq. (62), it can be easily found that

\begin{eqnarray}
  C_{2\alpha}-C_{\alpha}^{2}= C_{\alpha}^{2}
  (C_{\alpha}^{\frac{2\alpha}{(\alpha-1)}} - 1)
 \nonumber  \\
 C_{2\alpha-1}-C_{\alpha}^{2}= C_{\alpha}^{2}
  (C_{\alpha}^{\frac{2(\alpha-1)}{\alpha}} - 1).
\end{eqnarray} \\  
The Fisher information $I_{F}(p)$ is given by

\begin{eqnarray}
I_{F}(p)=
  \frac{\gamma^{2}}{(1-e^{-2\gamma t})}
 \nonumber  \\
 \times\Big( \frac{2e^{-2\gamma t}}{(e^{2\gamma t}-1)}
   + \frac{[\langle x(t)\rangle - s_{1}(t)]^{2}}{\sigma^{2}} \Big),
\end{eqnarray} \\  
and similarly for $I_{F}(q)$. Equation (68) indicates that temporal
Fisher information in this case is proportional (after a transient time)
to the square of the discrepancy between external signal and the average
particle position, which tries to track it. 

We can also compute entropy production rate for this system.
It is computed using the continuous formula \cite{hatano}:

\begin{eqnarray}
  \dot{S}_{i}= \int dx \; \frac{J_{i}(x,t)^{2}}{\gamma\sigma^{2}P_{i}(x,t)}.
\end{eqnarray} \\  
As a result we obtain for entropy production a similar formula as the one for the temporal
Fisher information. In particular, for the distribution $p(x,t)$ we have
$\dot{S}_{p}$ as

\begin{eqnarray}
\dot{S}_{p}= \gamma
\Big( \frac{e^{-2\gamma t}}{(e^{2\gamma t}-1)}
   + \frac{[\langle x(t)\rangle - s_{1}(t)]^{2}}{\sigma^{2}} \Big).
\end{eqnarray} \\  
This formula indicates that the higher the discrepancy between the
external signal and the average particle position, the larger the
entropy production rate.

Next, we address a question: how entropy production rate relates to the
power dissipated in the particle system? More general equation of motion
for our particle, if it had mass $m$, is given by:

\begin{eqnarray}
  m\ddot{x} = -\frac{\partial V_{i}(x,t)}{\partial x}  -k\dot{x}
  + k\sqrt{2\gamma\sigma^{2}}\eta(t).
\end{eqnarray}  
Both sides of this equation can be multiplied by the particle velocity
$\dot{x}$, which after a simple rearrangement yields

\begin{eqnarray}
  \frac{dE_{i}}{dt} =  \frac{\partial V_{i}(x,t)}{\partial t}  
  -k\dot{x}^{2}  + k\sqrt{2\gamma\sigma^{2}} \dot{x}\eta(t),
\end{eqnarray}  
where $E_{i}= \frac{1}{2}m\dot{x}^{2} + V_{i}(x,t)$ is the mechanical energy
of the system, and we used the relation
$\frac{dV_{i}(x,t)}{dt} = \frac{\partial V_{i}(x,t)}{\partial t}
+  \frac{\partial V_{i}(x,t)}{\partial x}\dot{x}$.  
Eq. (72) means that energy in the system is dissipated by three
different factors: by temporal decrease of the potential $V_{i}$, friction
proportional to $k\dot{x}^{2}$, and noise proportional to $\dot{x}\eta(t)$.
The average dissipated power, or energy rate, is
$\langle\frac{dE_{i}}{dt}\rangle$, which yields

\begin{eqnarray}
\langle\frac{dE_{i}}{dt}\rangle= 
  -k\gamma^{2}
  \Big( [\langle x(t)\rangle - s_{i}(t)]^{2}
% \nonumber  \\
  + \dot{s}_{i}[\langle x(t)\rangle - s_{i}(t)]
\nonumber  \\
  - \sigma^{2}e^{-2\gamma t}  \Big),
\end{eqnarray} \\  
where we used the Novikov theorem \cite{novikov} for the
average $\langle \dot{x}\eta(t)\rangle = \frac{1}{2}\sqrt{2\gamma\sigma^{2}}$.
Eq. (73) shows that the effect of noise on the energy change is negligible
after a transient time $\sim 1/\gamma$. The dominant contributions to
$\langle\frac{dE_{i}}{dt}\rangle$ are proportional to the discrepancy between
the external signal and the average particle position, which is similar
but not exactly the same as in the formula for the entropy production rate
(Eq. 70). The main difference between $\langle\frac{dE_{i}}{dt}\rangle$
and $\dot{S}_{p}$ is the term proportional to $\dot{s}_{i}$ (of any sign), which
characterizes energy flux (either positive or negative) between the
system and the environment. Finally, the form of Eq. (73) implies that
the average energy rate is bounded from above by 
$\langle\frac{dE_{i}}{dt}\rangle \le 
 k\gamma^{2}\big[ \sigma^{2}e^{-2\gamma t} + \frac{1}{4}(\dot{s}_{i})^{2} \big]$.

\subsubsection{Memory bistable systems.}

Here, we analyze a memory switch that can be driven by time dependent
external factors, and it is motivated by biophysics of small molecules.
We assume that this switch is a two state system, described by 
Markov dynamics. Examples of such bistable memory switches are proteins
(their activation and deactivation) and synapses in the brain
\cite{miller,petersen,karbowski2019}.

Let us consider a two state system with energies respectively $E_{1}$ and $E_{2}$
($E_{2} > E_{1}$), corresponding to states 1 and 2, that can be driven by a
time-dependent chemical potential. The system is at temperature $T$, which plays
the role of the noise, and there are stochastic jumps between states 1 and 2.
We assume that the system is initially ($t=0$) at equilibrium, and then a 
chemical potential $\mu(t)$ is turned on ($\mu(t) > 0$). This enables the system
to jump to the higher energy state 2, thus acquiring a new information above
a thermal background (the system learns). Our goal is to determine the rate of
this information gain, and its bounds in terms of physical quantities.
The master equation corresponding to this situation is:

\begin{eqnarray}
  \dot{p}_{1}= w_{12}p_{2} - w_{21}(t)p_{1},
  \nonumber  \\
  p_{2}= 1 - p_{1}
\end{eqnarray}
where the transition rate from the state 2 to 1 is $w_{12}=  e^{\beta\Delta E}/\tau$,
with $\tau$ the time scale for the jumps,
and the transition from the state 1 to 2 is driven by the chemical potential 
$\mu(t)$ as
$w_{21}=  e^{-\beta(\Delta E - \epsilon\mu(t))}/\tau$, with $\Delta E= E_{2}-E_{1} > 0$, 
$\beta= 1/(k_{B}T)$, and $\epsilon \ll 1$. The role of the chemical potentials is
to lower the energy barrier so that the jumps to the state 2 are more likely.
Initially, $t=0$, the system is at thermal equilibrium, and we have
$p_{1}(0)= e^{\beta\Delta E}/[2\cosh(\beta\Delta E)]$.

Eq. (74) can be solved exactly for an arbitrary form of $\mu(t)$. However, it is
convenient to work in the limit of small $\epsilon$ to get analytical expressions
for divergences and other physical quantities. We work to second order in $\epsilon$,
i.e., $p_{1}= p_{1}^{(0)} + \epsilon p_{1}^{(1)} + \epsilon^{2} p_{1}^{(2)} + O(\epsilon^{3})$,
and obtain  $p_{1}^{(0)}(t)= p_{1}(0)$, and

\begin{eqnarray}
  p_{1}^{(1)}(t)= - \frac{\beta}{\tau}e^{-w_{0}t}
  \Big[ p_{1}(0)e^{-\beta\Delta E} \int_{0}^{t}  ds \; \mu(s)
    \nonumber  \\
+ \frac{1}{\tau} \int_{0}^{t} ds \; e^{w_{0}s}\int_{s}^{t} ds' \; \mu(s') \Big]
\end{eqnarray} \\
where $w_{0}= (\frac{2}{\tau}) \cosh(\beta\Delta E)$. The form of $p_{1}^{(2)}$
does not appear in the expressions for the divergences and relevant physical
quantities to $\epsilon^{2}$ order, and thus it is not presented here explicitly.

The Chernoff $\alpha$-coefficient, between nonequilibrium probabilities
$p_{1}(t),p_{2}(t)$ and their equilibrium values $p_{1}(0),p_{2}(0)$,
is given by

\begin{eqnarray}
  C_{\alpha}(p(t)||p(0))= 1 + \epsilon^{2}\alpha(\alpha-1) \frac{(p_{1}^{(1)})^{2}}
  {2p_{1}^{(0)}p_{2}^{(0)}}  +  O(\epsilon^{3}),
\end{eqnarray}
which gives us immediately that Tsallis and Renyi divergences are identical
in this order:

\begin{eqnarray}
  T_{\alpha}(p(t)||p(0))= R_{\alpha}(p(t)||p(0))
  = \epsilon^{2}\frac{\alpha(p_{1}^{(1)})^{2}}
  {2p_{1}^{(0)}p_{2}^{(0)}}
 \nonumber  \\
  +  O(\epsilon^{3}).
\end{eqnarray}
They are both proportional to the square of nonequilibrium correction to the occupancy
probability.

The rate of Tsallis and Renyi divergences is

\begin{eqnarray}
\dot{T}_{\alpha}= \dot{R}_{\alpha}= - \frac{2\epsilon^{2}\alpha}{\tau} \cosh(\beta\Delta E)
  \Big[ e^{2\beta\Delta E} \beta\mu(t)p_{1}^{(1)}
 \nonumber  \\    
   + 4\cosh^{2}(\beta\Delta E)(p_{1}^{(1)})^{2} \Big]
  +  O(\epsilon^{3}),
\end{eqnarray}
and it is nonzero only if chemical potential is present, which corresponds to
the detailed balance breaking in the system. When $\dot{T}_{\alpha} > 0$
(or $\dot{R}_{\alpha} > 0$), then the system is gaining information, whereas
in the opposing case it is loosing information.

Similar to the previous example, the rates $\dot{T}_{\alpha}$ and
$\dot{R}_{\alpha}$ are quadratic in $p_{1}^{(1)}$, hence both of them 
are bounded from above by (for $\alpha > 0$)

\begin{eqnarray}
  \dot{T}_{\alpha}=  \dot{R}_{\alpha}
  \le \frac{\epsilon^{2}\alpha\beta^{2}e^{4\beta\Delta E}\mu(t)^{2}}
       {8\tau\cosh(\beta\Delta E)}     +  O(\epsilon^{3}).
\end{eqnarray}
Thus, the speed of divergence from the equilibrium is limited by the square
of the chemical potential and Boltzmann factors $e^{\beta\Delta E}$.

Alternatively, and more generally, one can use the bounds in Eqs. (13)
and (14), but for that some thermodynamic and information quantities
have to be determined first.

Temporal Fisher information is

\begin{eqnarray}
  I_{F}(p)= \frac{ \epsilon^{2}\big[ 4\cosh^{2}(\beta\Delta E) p_{1}^{(1)} + \beta\mu(t) \big]^{2}}
  {16\tau^{2}\cosh^{4}(\beta\Delta E)}
  +  O(\epsilon^{3}),
\end{eqnarray}
which means that the global speed of the system increases when chemical potential
is increasing.

Other physical quantities of interest are entropy production rates, $\dot{S}_{p}$,
and average activity $A_{p}= \langle \overline{w}\rangle$.
We find

\begin{eqnarray}
  \dot{S}_{p}= \frac{ \epsilon^{2}\big[ 4\cosh^{2}(\beta\Delta E) p_{1}^{(1)}
      + \beta\mu(t) \big]^{2}}
  {2\tau\cosh(\beta\Delta E)}
  +  O(\epsilon^{3}),
\end{eqnarray}
and for $A_{p} =  w_{21}p_{1} +  w_{12}p_{2}$ we have

\begin{eqnarray}
 A_{p} = \frac{1}{\tau \cosh(\beta\Delta E)} + 
  \frac{ \epsilon\big[ \beta\mu(t) - 2\sinh(2\beta\Delta E) p_{1}^{(1)}  
      \big]} {2\tau\cosh(\beta\Delta E)} 
 \nonumber  \\      
  +  O(\epsilon^{2}).
\end{eqnarray}
Equation (81) suggests that $\dot{S}_{p}$ grows quadratically with the chemical
potential and grows more nonlinearly with the energy barrier $\Delta E$.
On the contrary, the average activity $A_{p}$ decreases with $\Delta E$.

Note that, to the leading order, we have a simple relationship between
entropy production rate, Fisher information, and average activity:

\begin{eqnarray}
\dot{S}_{p}= \frac{8I_{F}(p)}{\tau^{2}A_{p}^{3}}
  +  O(\epsilon).
\end{eqnarray}
This equation suggests that the speed of learning $\sqrt{I_{F}(p)}$
(acquiring new information) is proportional (with different powers) to the
product of entropy production and average activity. Consequently,
it seems possible to maintain the speed of learning while simultaneously
decreasing dissipation and increasing internal activity.

\subsection{\label{sec:levell3} Applications in neuroscience.}

\subsubsection{Speed of gaining information during synaptic plasticity.}

It is believed that long-term information in real neural networks is
encoded collectively in the synaptic weights \cite{dayan,chklovskii,kasai,karbowski2023}.
Data from brain cortical networks suggest that synaptic weights are lognormally distributed,
characterized by heavy tails \cite{kasai,karbowski2023}. Such distributions seem
to be relatively stable during human development and adulthood \cite{karbowski2023}.

In what follows, we want to determine the bounds on the speed of gaining information
during synaptic learning. We assume that during synaptic plasticity, underlying learning
in neural circuits, synaptic weights change their mean and standard deviation values, but
they preserve their lognormal distributions. This assumption is consistent with the
stability of weights distribution during the lifetime \cite{karbowski2023}.

Let the initial probability density of synaptic weights $w$ (before learning at
time $t=0$) be $\rho_{0}(w)$, and during the learning phase (at times $t > 0$) be
$\rho(w,t)$. Thus, we have

\begin{eqnarray}
 \rho_{0}(w)= \frac{
    \exp\Big(- [\ln(w)-m_{0}]^{2}/2\sigma_{0}^{2} \Big) }
  {\sqrt{2\pi\sigma_{0}^{2}}w},
\end{eqnarray} \\
and
\begin{eqnarray}
 \rho(w,t)= \frac{
    \exp\Big(- [\ln(w)-m(t)]^{2}/2\sigma^{2}(t) \Big) }
  {\sqrt{2\pi\sigma^{2}(t)}w},
\end{eqnarray} \\
where $m_{0}$ and $\sigma_{0}^{2}$ are the mean and 
variance of logarithms of synaptic weights at $t= 0$, and
$m(t)$ and $\sigma^{2}(t)$ are the corresponding mean and
variance of logarithms of synaptic weights for $t > 0$.

The Chernoff $\alpha$-coefficient between $\rho(w,t)$ and $\rho_{0}(w)$
can be found as

\begin{eqnarray}
 C_{\alpha}(\rho||\rho_{0})= 
  \frac{\sigma_{0}^{\alpha}   \exp\Big( \frac{\alpha(\alpha-1)[m(t)-m_{0}]^{2}}
  {2[\alpha\sigma_{0}^{2} - (\alpha-1)\sigma^{2}(t)]} \Big) }
    {\sigma(t)^{\alpha-1} \sqrt{\alpha\sigma_{0}^{2} - (\alpha-1)\sigma^{2}(t)}},
\end{eqnarray} \\  
which is valid for $\alpha\sigma_{0}^{2} > (\alpha-1)\sigma^{2}(t)$, and
$C_{\alpha}(\rho||\rho_{0})= \infty$ for $\alpha\sigma_{0}^{2} \le (\alpha-1)\sigma^{2}(t)$.
This gives us, in the first case, the Renyi divergence 

\begin{eqnarray}
 R_{\alpha}(\rho||\rho_{0})= 
 \frac{\alpha(m(t)-m_{0})^{2}}
      {2[\alpha\sigma_{0}^{2} - (\alpha-1)\sigma^{2}(t)]}
      + \ln\frac{\sigma_{0}}{\sigma(t)}
 \nonumber  \\
 - \frac{\ln\big[1 + (\alpha-1)(1-\sigma^{2}(t)/\sigma_{0}^{2})\big]}
        {2(\alpha-1)},      
\end{eqnarray} \\  
and KL divergence

\begin{eqnarray}
D_{KL}(\rho||\rho_{0})= 
 \frac{(m(t)-m_{0})^{2}}{2\sigma_{0}^{2}}  + \ln\frac{\sigma_{0}}{\sigma(t)}
 \nonumber  \\
 - \frac{1}{2}\big[1-\frac{\sigma^{2}(t)}{\sigma_{0}^{2}}\big].      
\end{eqnarray} \\  
The KL divergence is the standard information gain during synaptic plasticity
\cite{cover,karbowski2024}, while the Renyi divergence is its generalization.
Their rates yield the speeds of gaining information. The rate of $R_{\alpha}$ is
somewhat complicated, but the rate of KL takes a simple form

\begin{eqnarray}
\frac{dD_{KL}}{dt}= 
 \frac{(m(t)-m_{0})\dot{m}(t)}{\sigma_{0}^{2}}  + \frac{\dot{\sigma}(t)}{\sigma(t)}
\Big[\frac{\sigma^{2}(t)}{\sigma_{0}^{2}} - 1 \Big].      
\end{eqnarray} \\  
The absolute values of both rates are bounded by the inequalities in Eqs. (14),
(15) and (17), with the temporal Fisher information:

\begin{eqnarray}
I_{F}(\rho)= 
 \frac{2\dot{\sigma}(t)^{2} + \dot{m}(t)^{2}}{\sigma^{2}(t)}.  
\end{eqnarray} \\  
This means that the speeds of gaining information during synaptic plasticity,
while learning, are limited mostly by the speeds of changing the two
parameters characterizing means and variances of synaptic weights.

An interesting question is when the bound in Eq. (15) for the plastic
synapses is saturated. To answer this question, we have to first determine
$\langle \ln^{2}\frac{\rho(w,t)}{\rho_{0}(w)} \rangle_{\rho}$. It can be
easily found as:

\begin{eqnarray}
\langle \ln^{2}\frac{\rho(w,t)}{\rho_{0}(w)} \rangle_{\rho}=
D_{KL}(\rho||\rho_{0})^{2} +
\frac{[\sigma^{2}(t) - \sigma_{0}^{2}]^{2}}{2\sigma_{0}^{4}}      
 \nonumber  \\
+ \frac{\sigma^{2}(t)}{\sigma_{0}^{4}}[m(t)-m_{0}]^{2}.
\end{eqnarray} \\
With this, all the terms in Eq. (15) are given explicitly.
After some arrangements, we find that the general inequality in
Eq. (15) is equivalent to the following specific inequality:

\begin{eqnarray}
0 \le 
[2\sigma(t)\dot{\sigma}(t)(m(t)-m_{0}) -  \dot{m}(t)(\sigma^{2}(t)-\sigma_{0}^{2})]^{2}.
\end{eqnarray} \\  
This implies that the bound in Eq. (15) is saturated if the right hand side
of Eq. (92) is 0, which is satisfied when the ratio
$(m(t)-m_{0})/(\sigma^{2}(t)-\sigma_{0}^{2})= const$. In other words, the mean
and variance of the logarithm of synaptic weights must change in a coordinated manner,
which is a very restrictive condition on the stochastic dynamics of synapses.

\subsubsection{Predictive inference.}

The bounds presented in Eqs. (12-15) could be also used in predicting the future
behavior of a stochastic dynamical system. In particular, the brain neural networks
have to often make predictions about some external signal, which is somehow important
for the organism possessing that brain \cite{still,bialek2001,lang,palmer,still2014}.
Let the external time-dependent sensory signal be $x_{t}$. Neurons in the sensory
cortex of the brain try to predict the value of the signal $x_{t'}$ in future times
$t' > t$ \cite{bialek2001,palmer,still2014}, using some internal dynamical variable
$m_{t}$ that relates to neural and synaptic activities. One can think about $m_{t}$ as
some sort of ``memory'' variable, which keeps the information about the past of the
signal $x_{t}$ up to the time $t$, in a compressed manner. In the simplest situation,
neural activity $m_{t}$ tries to predict the signal at the nearest future, i.e.,
to estimate the value $x_{t+\Delta t}$, where $\Delta t$ is small. The key in this
estimate are two conditional probabilities: $p(x_{t+\Delta t}|x_{t})$ and
$p(x_{t+\Delta t}|m_{t})$. The former is the probability of the jump
in the signal value from time $t$ to time $t+\Delta t$, and the latter is
the probability of the signal at time $t+\Delta t$ given the value $m_{t}$
of the memory variable at time $t$. Thus, the first conditional probability
describes a natural temporal evolution of the external signal, whereas the
second is the estimate of this evolution given the knowledge of neural
activity $m_{t}$.

The goodness of predictability can be quantified by KL divergence between
actual external dynamics $p(x_{t+\Delta t}|x_{t})$  and its prediction
$p(x_{t+\Delta t}|m_{t})$, i.e. \cite{still2014}

\begin{eqnarray}
  D_{KL}(p(x_{t+\Delta t}|x_{t})||p(x_{t+\Delta t}|m_{t})=
 \nonumber  \\
  \int \; dx_{t+\Delta t} \; p(x_{t+\Delta t}|x_{t})
 \ln\frac{p(x_{t+\Delta t}|x_{t})}{p(x_{t+\Delta t}|m_{t})}.
\end{eqnarray} \\  
The smaller the value of $D_{KL}$, the better the memory variable predicts
the external dynamics. The rate of $D_{KL}$ measures how fast the prediction
can deteriorate. In this sense, Eqs. (15) and (17) provide bounds on the
speed of predictability degradation. These bounds are determined to a large
extent by the temporal Fisher informations: $I_{F}(p(x_{t+\Delta t}|x_{t}))$,
which gives the square of the speed of transitions in the external signal, 
and $I_{F}(p(x_{t+\Delta t}|m_{t}))$, which yields the speed of external
dynamics given the instantaneous value of the memory variable.

Below we analyze a specific example of predictive inference, in which
one can obtain explicit formulas for all relevant variables appearing
in Eqs. (15) and (17). Let us consider the following dynamics for $x_{t}$
and $m_{t}$:

\begin{eqnarray}
  \dot{x}= -\gamma(x-\lambda_{t}) + \sqrt{2\gamma\sigma^{2}}\eta(t),
\end{eqnarray}  
\begin{eqnarray}
  \dot{m}= -\gamma_{m}(m-x) + \sqrt{2\gamma_{m}\sigma_{m}^{2}}\eta_{m}(t),
\end{eqnarray}  
where $\lambda_{t}$ is some external dynamical variable governing the trajectory
$x_{t}$, $\gamma$ and $\gamma_{m}$ are inverses of time constants for the external
and internal systems, $\sigma$ and $\sigma_{m}$ are standard deviations of
the Gaussian noise terms $\eta$ and $\eta_{m}$, similar as in Eq. (58).
For simplicity, let us consider the case $\gamma_{m}/\gamma \gg 1$, which
corresponds to the situation in which the dynamics of internal variable $m_{t}$
is much faster than the external $x_{t}$. The discretized version of Eqs. (94)
and (95) in this limit has the following forms

\begin{eqnarray}
  x_{t+\Delta t}= x_{t} -\gamma(x_{t}-\lambda_{t})\Delta t + \sqrt{2\gamma\sigma^{2}}
  \eta_{t}\Delta t,
\end{eqnarray}  
\begin{eqnarray}
 x_{t} \approx  m_{t} - \sqrt{\frac{2\sigma_{m}^{2}}{\gamma_{m}}}\eta_{m,t}.
\end{eqnarray}  
Note that the memory variable $m_{t}$ differs in this limit from the actual external
variable $x_{t}$ only by appropriately rescaled noise term.

The relevant conditional probabilities can be found from Eqs. (96) and (97) as
\cite{risken}

\begin{eqnarray}
  p(x_{t+\Delta t}|x_{t})= \frac{\exp\Big(
    -\frac{[x_{t+\Delta t}-x_{t}+\gamma(x_{t}-\lambda_{t})\Delta t]^{2}}
   {4\gamma\sigma^{2}\Delta t} \Big) }
{\sqrt{4\pi\gamma\sigma^{2}\Delta t}}   
\end{eqnarray}  
\begin{eqnarray}
  p(x_{t}|m_{t})= 
  \sqrt{\frac{\gamma_{m}\Delta t}{4\pi\sigma_{m}^{2}}}
    \exp\Big(- \frac{\gamma_{m}\Delta t}{4\sigma_{m}^{2}} [x_{t}-m_{t}]^{2}\Big),
\end{eqnarray}  
where Eq. (98) is valid for small time interval $\Delta t$.
The remaining conditional probability $p(x_{t+\Delta t}|m_{t})$ of interest is found
from the relation

\begin{eqnarray}
p(x_{t+\Delta t}|m_{t})=
  \int \; dx_{t} \; p(x_{t+\Delta t}|x_{t}) p(x_{t}|m_{t}),
\end{eqnarray} \\  
which after a straightforward calculation yields

\begin{eqnarray}
  p(x_{t+\Delta t}|m_{t})= \sqrt{\frac{\gamma_{m}\Delta t}
{4\pi(\gamma\gamma_{m}\sigma^{2}(\Delta t)^{2} + \sigma_{m}^{2})} }  
 \nonumber  \\
\times \exp\Big(
  -\frac{\gamma_{m}\Delta t[x_{t+\Delta t}-m_{t}+\gamma(x_{t}-\lambda_{t})\Delta t]^{2}}
   {4(\gamma\gamma_{m}\sigma^{2}(\Delta t)^{2} + \sigma_{m}^{2})} \Big).
\end{eqnarray}  
Note that $p(x_{t+\Delta t}|m_{t})$ and $p(x_{t+\Delta t}|x_{t})$ become
identical, with substitution $m_{t} \leftrightarrow x_{t}$, if
$\gamma_{m} \mapsto \infty$.

Eqs. (93), (98) and (101) allow us to find the KL divergence
$D_{KL}(p(x_{t+\Delta t}|x_{t})||p(x_{t+\Delta t}|m_{t})$ as

\begin{eqnarray}
  D_{KL}(p(x_{t+\Delta t}|x_{t})||p(x_{t+\Delta t}|m_{t}) =
  \frac{1}{2} \ln\Big( 1 + \frac{\sigma_{m}^{2}}{\gamma\gamma_{m}\sigma^{2}(\Delta t)^{2}} \Big)
 \nonumber  \\
 + \frac{ \gamma_{m}\Delta t(x_{t}-m_{t})^{2} - 2\sigma_{m}^{2}}
 {4[\gamma\gamma_{m}\sigma^{2}(\Delta t)^{2} + \sigma_{m}^{2}]},
  \nonumber  \\
\end{eqnarray}  
and its temporal derivative as

\begin{eqnarray}
 \frac{dD_{KL}}{dt}= \frac{ \gamma_{m}\Delta t(x_{t}-m_{t})\dot{x}_{t}}
 {2[\gamma\gamma_{m}\sigma^{2}(\Delta t)^{2} + \sigma_{m}^{2}]}.
\end{eqnarray}  
The rate of KL divergence in this case is proportional to the speed of change
in the external variable $x_{t}$ (it does not depend on the speed of $m_{t}$
because $\dot{m}_{t}\approx 0$ in the limit $\gamma_{m}/\gamma \gg 1$), and
to the instant difference between $x_{t}$ and $m_{t}$. Consequently, the inference
of the external signal improves if the rate of $D_{KL}$ divergence decreases, which
takes place when $\dot{x}_{t}$ and $(x_{t}-m_{t})$ have the opposite signs.
For example, if the external signal slows down ($\dot{x}_{t} < 0$), then the
prediction internal variable $m_{t}$ should be smaller than the signal $x_{t}$
to get closer to it in the next instant of time, i.e., to improve the prediction.

Eq. (103) is the exact form of the speed of divergence between the true
external dynamics $x_{t}$ and its estimate using internal variable $m_{t}$.
Alternatively, one can provide the bounds on $dD_{KL}/dt$ using Eqs. (15)
and (17). For this, one needs to determine the temporal Fisher informations,
defined as

\begin{eqnarray}
I_{F}(p(x_{t+\Delta t}|x_{t}))=
  \int \; dx_{t+\Delta t} \; p(x_{t+\Delta t}|x_{t})
 \nonumber  \\
\times \Big[\frac{\dot{p}(x_{t+\Delta t}|x_{t})}{p(x_{t+\Delta t}|x_{t})}\Big]^{2}
\end{eqnarray} \\  
and
\begin{eqnarray}
I_{F}(p(x_{t+\Delta t}|m_{t}))=
  \int \; dx_{t+\Delta t} \; p(x_{t+\Delta t}|m_{t})
 \nonumber  \\
\times \Big[\frac{\dot{p}(x_{t+\Delta t}|m_{t})}{p(x_{t+\Delta t}|m_{t})}\Big]^{2},
\end{eqnarray} \\
which describe the speed of transitions in the external signal and
the speed of predictions of that signal, respectively.
After some simple algebra, one can find both Fisher informations as

\begin{eqnarray}
I_{F}(p(x_{t+\Delta t}|x_{t}))=
 \frac{\gamma\Delta t(\dot{x}_{t}-\dot{\lambda}_{t})^{2}}
 {2\sigma^{2}},
\end{eqnarray}  
and
\begin{eqnarray}
I_{F}(p(x_{t+\Delta t}|m_{t}))=
 \frac{\gamma_{m}\Delta t(\dot{x}_{t})^{2}}
 {2[\gamma\gamma_{m}\sigma^{2}(\Delta t)^{2} + \sigma_{m}^{2}]},
\end{eqnarray}  
and both of them depend on the speed of the external signal $\dot{x}_{t}$.
Note that for $\gamma_{m}/\gamma \gg 1$ the second Fisher information
$I_{F}(p(x_{t+\Delta t}|m_{t}))$ dominates over the first one, since
generally $\gamma\Delta t \ll 1$. This means that the speed of prediction
is much larger than the speed of the external signal, which is consistent
with our initial assumption.

\section{\label{sec:level3}  Bounds on the rate of mutual information.}

Mutual information $I(x,y)$ between two stochastic variables $x, y$ with
joint probability $p_{xy}(x,y)$, and marginal probabilities $p_{x}(x)$, $p_{y}(y)$,
can be defined as KL divergence between probabilities $p$ and $q$  given by
$p= p_{xy}(x,y)$, and $q= p_{x}(x)p_{y}(y)$. More precisely,
$I(x,y)= D_{KL}(p_{xy}||p_{x}p_{y})$.

\subsection{\label{sec:levell3} General kinematic bounds.}

General kinematic bound on the rate of mutual information $I(x,y)$,
irrespective of the type of systems dynamics, follows from Eq. (15) and (17),
and takes the form

\begin{eqnarray}
 | \frac{dI(x,y)}{dt}| \le
  \sqrt{I_{F}(p_{xy})}\sqrt{\langle \ln^{2}\frac{p_{xy}}{p_{x}p_{y}}\rangle -I(x,y)^{2}}
  \nonumber  \\
  +  \sqrt{I_{F}(p_{x})+I_{F}(p_{y})}\sqrt{\langle \frac{p_{xy}}{p_{x}p_{y}}\rangle - 1}
\nonumber  \\ 
\le  \sqrt{I_{F}(p_{xy})}\sqrt{C_{2}^{xy} -1 - I(x,y)^{2}}
  \nonumber  \\
  +  \sqrt{I_{F}(p_{x})+I_{F}(p_{y})}\sqrt{C_{2}^{xy} - 1},
\end{eqnarray}
where $C_{2}^{xy}$ is the Chernoff coefficient, i.e.,
$C_{2}^{xy}= \langle \frac{p_{xy}}{p_{x}p_{y}}\rangle$, and averaging is done
with respect to the joint probability $p_{xy}$. We used the fact that Fisher
information of the product of probabilities decomposes into the sum, i.e.,
$I_{F}(p_{x}p_{y})= I_{F}(p_{x}) + I_{F}(p_{y})$. The upper bound on
$dI/dt$ is thus constrained by the global dynamical rates of the whole
system $x,y$ and of the subsystems $x$ and $y$ (it is called here bound
$B_{I}1$). These rates are appropriately rescaled by the degrees of mutual
correlations between variables $x$ and $y$.

\subsection{\label{sec:levell3} Examples: weakly correlated systems.}

In this section we check the quality of the bound represented by Eq. (108)
for two examples of weakly coupled systems X and Y, with continuous
state variables $x$ and $y$, respectively.

\subsubsection{ Bivariate Gaussian distribution.}

Consider the system (X,Y) of weakly correlated variables $x$ and $y$
with Gaussian joint probability density

\begin{eqnarray}
  p_{xy}(x,y)=  \frac{\exp\big(-[\overline{x},\overline{y}]\Sigma^{-1}
  [\overline{x},\overline{y}]^{T}\big) }
  {2\pi\sqrt{\det(\Sigma)}},
\end{eqnarray}
where $[\overline{x},\overline{y}]$ is two dimensional vector
with the components $\overline{x}= x - \mu_{x}$, and $\overline{y}= y - \mu_{y}$,
where $\mu_{x}$ and $\mu_{y}$ denote mean values of $x$ and $y$.
The symbol $\Sigma$ is $2\times 2$ covariance matrix with elements:
$\Sigma_{11}= \sigma_{x}^{2}$, $\Sigma_{22}= \sigma_{y}^{2}$,
and $\Sigma_{12}= \Sigma_{21}= \epsilon r_{0}(t)\sigma_{x}\sigma_{y}$,
where $\epsilon \ll 1$. We assume that only the correlation coefficient
$r_{0}$ is time dependent.

Consequently, marginal distributions
$p_{x}(x)= \frac{\exp(-\overline{x}^{2}/2\sigma_{x}^{2})}{\sqrt{2\pi\sigma_{x}^{2}}}$
  and
$p_{y}(y)= \frac{\exp(-\overline{y}^{2}/2\sigma_{y}^{2})}{\sqrt{2\pi\sigma_{y}^{2}}}$
  are time independent, which implies that
temporal Fisher information for these distributions
$I_{F}(p_{x})= I_{F}(p_{y})= 0$.  

For small $\epsilon$ the joint density (109) separates into
  a product of the marginal distributions as
  
\begin{eqnarray}
  p_{xy}= p_{x}p_{y}\Big( 1 + \frac{\epsilon r_{0}}{\sigma_{x}\sigma_{y}}
  \overline{x}\overline{y}
%  \nonumber  \\  
  + \frac{1}{2}\big(\frac{\epsilon r_{0}}{\sigma_{x}\sigma_{y}}\big)^{2}
  \nonumber  \\  
  \times \big[ 2(\overline{x}\overline{y})^{2} + 
    (\sigma_{x}\sigma_{y})^{2} - \sigma_{y}^{2}\overline{x}^{2} 
    - \sigma_{x}^{2}\overline{y}^{2} \big] + O(\epsilon^{3}) \Big),
\end{eqnarray}
which allows us to compute all the variables in Eq. (108) to the
lowest order in $\epsilon$.

Thus, mutual information $I(x,y)$ is \cite{cover}

\begin{eqnarray}
  I(x,y)= -\frac{1}{2}\ln(1-\epsilon^{2}r_{0}(t)^{2})
  = \frac{1}{2}\epsilon^{2}r_{0}(t)^{2}  + O\big(\epsilon^{3}\big),
  \nonumber  \\  
\end{eqnarray}
and temporal Fisher information for the joint density in Eq. (109)
is

\begin{eqnarray}
  I_{F}(p_{xy})= \langle \Big(\frac{\dot{p}_{xy}}{p_{x}p_{y}}\Big)^{2}\rangle =
  \epsilon^{2} \dot{r}_{0}(t)^{2}  + O\big(\epsilon^{3}\big),
\end{eqnarray}
where $\dot{r}_{0}(t)= dr_{0}/dt$. Moreover, we have

\begin{eqnarray}
  \langle \ln^{2}\big(\frac{p_{xy}}{p_{x}p_{y}}\big) \rangle =
  \epsilon^{2}r_{0}(t)^{2} + O\big(\epsilon^{3}\big).
\end{eqnarray}

With Eqs. (111-113) we can find the left and right hand sides of Eq. (108).
It turns out that both sides of this equation are equal to each other in
the lowest order of $\epsilon$, with the values 
$\epsilon^{2}r_{0}(t)|\dot{r}_{0}| + O\big(\epsilon^{3}\big)$ each.
This implies that for weakly correlated Gaussian variables the
bound on the rate of mutual information in Eq. (108) is saturated,
if $\sigma_{x}$ and $\sigma_{y}$ are time independent. However, the situation
is more complicated if the variances of $x$ and $y$ are time dependent.
In this case, much depends on how big are the rates of $\sigma_{x}$ and $\sigma_{y}$.
If they are small, of the order $\sim \epsilon$, then the bound in Eq. (108)
is close to saturation, but if they are large $\sim O(1)$ then the
left hand side is much smaller (by the factor proportional to $\epsilon$)
than the right hand side. This follows from the observation that in this
case all Fisher informations, $I_{F}(p_{xy})$, $I_{F}(p_{x})$ and $I_{F}(p_{x})$
would be of order $O(1)$.

\subsubsection{ Bivariate non-Gaussian distribution.}

In this case we choose the joint probability density (normalized)
for the weakly correlated system (X,Y) in the form:

\begin{eqnarray}
  p_{xy}(x,y)= \frac{(\kappa_{1}\kappa_{2})^{2}[1+\epsilon r(t)xy]}
  {\kappa_{1}\kappa_{2}+\epsilon r(t)} 
  e^{-(\kappa_{1}x +\kappa_{2}y)},
\end{eqnarray}
where $\kappa_{1}, \kappa_{2}$ are some positive constants, $\epsilon \ll 1$,
and $r(t)$ is time dependent positive parameter
characterizing the degree of correlations between X and Y.
Note that for $\epsilon= 0$ the systems X and Y become decoupled.
The marginal probability densities $p_{x}$, $p_{y}$ are time dependent
and read

\begin{eqnarray}
  p_{x}(x)= \frac{\kappa_{1}^{2}\kappa_{2}[1+\epsilon r(t)\frac{x}{\kappa_{2}}]}
  {\kappa_{1}\kappa_{2}+\epsilon r(t)} 
  e^{-\kappa_{1}x},
  \nonumber  \\
  p_{y}(y)= \frac{\kappa_{1}\kappa_{2}^{2}[1+\epsilon r(t)\frac{y}{\kappa_{1}}]}
  {\kappa_{1}\kappa_{2}+\epsilon r(t)} 
  e^{-\kappa_{2}y}.
\end{eqnarray}

As in the Gaussian case, all the relevant variables present in Eq. (108) can be
computed analytically for this case as a series expansion in the small parameter
$\epsilon$ (see Appendix D). Consequently, mutual information $I(x,y)$ to the lowest
order in $\epsilon$ is 

\begin{eqnarray}
  I(x,y)= \frac{\epsilon^{2}}{2} \Big( \frac{r(t)}{\kappa_{1}\kappa_{2}} \Big)^{2}
  + O\big(\epsilon^{3}\big).
\end{eqnarray}

Temporal Fisher informations $I_{F}(p_{xy})$, and $I_{F}(p_{x})$, $I_{F}(p_{y})$
take the forms:

\begin{eqnarray}
  I_{F}(p_{xy})= 3\epsilon^{2} \Big( \frac{\dot{r}(t)}{\kappa_{1}\kappa_{2}} \Big)^{2}
  + O\big(\epsilon^{3}\big),
  \nonumber  \\
  I_{F}(p_{x})= I_{F}(p_{y}) = \epsilon^{2} \Big( \frac{\dot{r}(t)}
  {\kappa_{1}\kappa_{2}} \Big)^{2}
  + O\big(\epsilon^{4}\big),
\end{eqnarray}
where $\dot{r}(t)= dr/dt$. Additionally, we have

\begin{eqnarray}
  \langle \ln^{2}\big(\frac{p_{xy}}{p_{x}p_{y}}\big) \rangle =
  \epsilon^{2} \Big( \frac{r(t)}{\kappa_{1}\kappa_{2}} \Big)^{2}
 + O\big(\epsilon^{3}\big),
 \nonumber  \\
  \langle \frac{p_{xy}}{p_{x}p_{y}} \rangle =
  1 + \epsilon^{2} \Big( \frac{r(t)}{\kappa_{1}\kappa_{2}} \Big)^{2}
 + O\big(\epsilon^{3}\big).
\end{eqnarray}

Insertion of Eqs. (116-118) to Eq. (108) gives us the left hand
side, $|dI(x,y)/dt|$, equal to 
$ \big(\frac{\epsilon}{\kappa_{1}\kappa_{2}}\big)^{2} r(t)|\dot{r}(t)|$,
whereas both right hand sides are identical to the lowest order and
equal to 
$(\sqrt{2}+\sqrt{3}) \big(\frac{\epsilon}{\kappa_{1}\kappa_{2}}\big)^{2}
r(t)|\dot{r}(t)|$. This means that, for weakly correlated systems,
the bound on the rate of mutual information in Eq. (108) 
is greater than its actual absolute value by a factor of 3.1,
which is a worse performance than for the Gaussian case.
Thus, comparing the two examples, it is clear that the degree of
accuracy in estimation of $|dI(x,y)/dt|$ is model dependent.

\subsection{\label{sec:levell3} Kinematic-thermodynamic bounds for Markov
  processes.}

In this section we provide upper bounds on the rates of mutual information
for more restricted dynamics of composite Markov systems with interacting
systems X and Y. First we discuss dynamics of such systems and then
give the bounds on the rate of mutual information between X and Y.

\subsubsection{Master equation for composite systems.}

For the composite system (X,Y) containing two interacting subsystems X and Y
we can formulate Master equation similar to Eq. (22)

\begin{eqnarray}
  \dot{p}_{xy}= \sum_{x',y'} \Big({\bf W}(xy|x'y')p_{x'y'}
  - {\bf W}(x'y'|xy)p_{xy}\Big),
%  \nonumber  \\
\end{eqnarray}         
where $p_{xy}$ is the joint probability that the system (X,Y) is in the
state $(x,y)$, and ${\bf W}(x'y'|xy)$ is the global transition rate
from state $(x,y)$ to state $(x',y')$.

Average activity $A_{xy}$ of the whole system (XY) corresponds to the
first moment of the total escape rate and is given by
(\cite{koyuk,vo}; compare Eq. 25)

\begin{eqnarray}
  A_{xy} \equiv  \langle \overline{{\bf W}}\rangle_{xy}=
  \sum_{x,y}  \overline{{\bf W}}(xy)p_{xy},
\end{eqnarray}
where the total escape rate from state $(xy)$ is 
$\overline{{\bf W}}(xy)= \sum_{x',y'} {\bf W}(x'y'|xy)$.

Second moment of the total escape rate of composite system is defined as

\begin{eqnarray}
\langle {\overline{\bf W}^{2}}\rangle_{xy}= \sum_{x,y} \overline{{\bf W}}(xy)^{2}p_{xy},
\end{eqnarray}
and the system entropy production rate $\dot{S}_{xy}$ is 

\begin{eqnarray}
  \dot{S}_{xy}= \frac{1}{2} \sum_{xy,x'y'} \big({\bf W}(xy|x'y')p_{x'y'}
  - {\bf W}(x'y'|xy)p_{xy}\big)
 \nonumber  \\  
 \times \ln\frac{{\bf W}(xy|x'y')p_{x'y'}}{{\bf W}(x'y'|xy)p_{xy}}.
  \nonumber  \\  
\end{eqnarray}

In a particular case when the joint probability is separable, i.e.,
$p_{xy}= p_{x}p_{y}$, we denote the above quantities as:
 $A^{(0)}_{xy} \equiv  \langle \overline{{\bf W}}\rangle_{xy}^{(0)}$
for average activity, $\langle {\overline{\bf W}^{2}}\rangle^{(0)}_{xy}$
for the second moment of total escape rate, and
$\dot{S}^{(0)}_{xy}$ as entropy production rate.

For the so-called bipartite systems \cite{hartich,diana,horowitz}
the global transition rates ${\bf W}(x'y'|xy)$ in Eq. (119) can be
written in a more explicit form as
${\bf W}(x'y'|xy)= w_{x'x}^{y}\delta_{yy'} + w_{y'y}^{x}\delta_{xx'}$.
In this formula $w_{x'x}^{y}, w_{y'y}^{x}$ are transition rates in the
subsystems X and Y respectively, which depend on the actual state in
the neighboring system. Here, one considers only single transitions
in either of the subsystems ($x \mapsto x'$ or $y \mapsto y'$),
neglecting double simultaneous transitions, i.e. $(x,y) \mapsto (x',y')$,
as they are less likely. With the above decomposition of 
${\bf W}(x'y'|xy)$ one can write the master equation (119) as

\begin{eqnarray*}
  \dot{p}_{xy}= \sum_{x'} w^{y}_{xx'}p_{x'y} + \sum_{y'} w^{x}_{yy'}p_{xy'}
 \nonumber  \\  
  - (\sum_{x'} w^{y}_{x'x} + \sum_{y'} w^{x}_{y'y})p_{xy}.
\end{eqnarray*}         
Moreover, also $A_{xy}$, $\langle {\overline{\bf W}^{2}}\rangle_{xy}$,
and $\dot{S}_{xy}$ can be expressed in terms of local transition rates
$w_{x'x}^{y}, w_{y'y}^{x}$, which may be useful \cite{hartich,diana,horowitz}
(see also below).

\subsubsection{The rates of mutual information for Markov processes.}

For the Markov dynamics represented by the Master equation (119),
the thermodynamic-kinematic bound on the rate of mutual information
between variables $x$ and $y$ can be deduced from Eq. (40), and reads 

\begin{widetext}
\begin{eqnarray}
 | \frac{dI(x,y)}{dt}|  \le
  \Big([I_{F}(p_{x})+I_{F}(p_{y})] \sqrt{A^{(0)}_{xy}\Psi^{(0)}_{xy}}
  \big[ C_{3}^{xy} - C_{2}^{xy} \big]\Big)^{1/3}
  \nonumber  \\    
+ \Big(I_{F}(p_{xy})\sqrt{A_{xy}\Psi_{xy}}
\big[ e^{-3I/2} C_{5/2}^{xy}
  - e^{-I/2} C_{3/2}^{xy}
  - e^{I/2} C_{1/2}^{xy}
  + e^{3I/2} C_{-1/2}^{xy}
  \big] \Big)^{1/3},
%  \nonumber  \\    
\end{eqnarray}         
\end{widetext}
where the Chernoff coefficient
$C_{\alpha}^{xy}= \langle \big(\frac{p_{xy}}{p_{x}p_{y}}\big)^{\alpha-1}\rangle$,
and averaging is done with respect to $p_{xy}$. Additionally, the symbols
$\Psi_{xy}$ and $\Psi^{(0)}_{xy}$ are analogs of Eq. (35) with the joint
probability $p_{xy}$ and its separable variant $p_{xy}= p_{x}p_{y}$, respectively.

An alternative kinematic-thermodynamic bound on the rate of mutual
information, coming from Eqs. (45) and (46) is

\begin{widetext}
\begin{eqnarray}
 |\frac{dI(x,y)}{dt}| \le
 \sqrt{\Psi_{xy}/2}
\Big(\sqrt{I_{F}(p_{xy})} + 2\sqrt{\langle\overline{\bf W}^{2}\rangle_{xy}}\Big)^{1/2}
\big( e^{-2I} C_{3}^{xy} - 4e^{-I} C_{2}^{xy} 
  - 4e^{I} + e^{2I} C_{-1}^{xy} + 6 \big)^{1/4}
   \nonumber  \\
+ \sqrt{\Psi^{(0)}_{xy}/2}
\Big(\sqrt{I_{F}(p_{x})+I_{F}(p_{y})}
+ 2\sqrt{\langle\overline{\bf W}^{2}\rangle^{(0)}_{xy}}\Big)^{1/2}
\Big(C_{4}^{xy} - 4C_{3}^{xy}  +  6C_{2}^{xy} -3\Big)^{1/4}.
\end{eqnarray}
\end{widetext}

The inequalities (123) and (124), called here bounds $B_{I}2, B_{I}3$
and $B_{I}4$, $B_{I}5$ respectively, imply that for Markov dynamical
physical systems the rates at which we can gain information
about one variable (X) by observing another correlated variable
(Y) can be restricted in several ways, but the speeds of global
and internal dynamics always appear in those limitations. This suggests
that the speeds of information transfer between two systems are limited
primary by global and local speeds of system transformations. Since the
speed of local dynamics is related to activity $A$ and entropy production rate,
this also mean that the rate of MI is constrained by thermodynamics, which is
in line with previous conclusions \cite{barato}.
Moreover, out of the three bounds on $dI(x,y)/dt$ represented by Eqs. (108), (123)
and (124), the most accurate is the bound given by Eq. (108). This follows from
the fact that this bound is the most general, as it is independent of the type of
system dynamics (either Markovian or non-Markovian). This is also supported
by the results in Figs. 1 and 2, where the bound B1 (corresponding to Eq. (108))
is very close the actual value of $|dC_{\alpha}/dt|$.

It is important to note that the bounds on the rate of MI given by Eqs. (108),
(123) and (124) are a new type of bounds, and they are different from other
existing bounds on mutual information or its rate \cite{barato,paninski,kraskov}.
The derived here inequalities for the rate of MI are collected in Table 2.

\begin{widetext}
\begin{table}
\begin{center}
\caption{Summary of the inequalities for the rates of MI.}
\begin{tabular}{|l c |}
\hline

Bound    &       Equation        \\  
type     &                     \\

\hline

Kinematic:   &                   \\
$B_{I}1$  &
$|\dot{I}_{xy}| \le \sqrt{I_{F}(p_{xy})}\sqrt{C_{2}^{xy} -1 - I_{xy}^{2}}
  +  \sqrt{I_{F}(p_{x})+I_{F}(p_{y})}\sqrt{C_{2}^{xy} - 1}$  \\
 &    \\

\hline

    &                   \\
Kinematic-thermodynamic:  &                   \\
$B_{I}2$, $B_{I}3$      &
 $|\dot{I}_{xy}|  \le
  \Big([I_{F}(p_{x})+I_{F}(p_{y})] \sqrt{A^{(0)}_{xy}\Psi^{(0)}_{xy}}
  \big[ C_{3}^{xy} - C_{2}^{xy} \big]\Big)^{1/3}$  \\ 
&    \\
&  $ + \Big(I_{F}(p_{xy})\sqrt{A_{xy}\Psi_{xy}}
\big[ e^{-3I_{xy}/2} C_{5/2}^{xy}  - e^{-I_{xy}/2} C_{3/2}^{xy}
  - e^{I_{xy}/2} C_{1/2}^{xy}  + e^{3I_{xy}/2} C_{-1/2}^{xy} \big] \Big)^{1/3}$  \\

&    \\

    &                   \\
$B_{I}4$, $B_{I}5$     &
$|\dot{I}_{xy}| \le
\sqrt{\Psi^{(0)}_{xy}/2}\Big(\sqrt{I_{F}(p_{x})+I_{F}(p_{y})}
+ 2\sqrt{\langle\overline{\bf W}^{2}\rangle^{(0)}_{xy}}\Big)^{1/2}
\Big(C_{4}^{xy} - 4C_{3}^{xy}  +  6C_{2}^{xy} -3\Big)^{1/4} $   \\
&    \\
&  $ + \sqrt{\Psi_{xy}/2}
\Big(\sqrt{I_{F}(p_{xy})} + 2\sqrt{\langle\overline{\bf W}^{2}\rangle_{xy}}\Big)^{1/2}
\big( e^{-2I_{xy}} C_{3}^{xy} - 4e^{-I_{xy}} C_{2}^{xy} 
  - 4e^{I_{xy}} + e^{2I_{xy}} C_{-1}^{xy} + 6 \big)^{1/4} $ \\

\hline
\end{tabular}
\end{center}

\end{table}
\end{widetext}

\section{\label{sec:level3} Application of inequalities for the rates of MI:
Bipartite sensing and Landauer limit}

For bipartite systems, with X describing external variable and Y
internal variable, one can decompose the rate of mutual information,
$\dot{I}_{xy}\equiv dI(x,y)/dt$, between X and Y into the so-called information
flows $\dot{I}_{x}$ and $\dot{I}_{y}$ as \cite{allahverdyan,horowitz}

\begin{eqnarray}
  \dot{I}_{xy}= \dot{I}_{x} + \dot{I}_{y},
\end{eqnarray}         
where $\dot{I}_{x}= \big[I(x_{t+dt},y_{t})-I(x_{t},y_{t})\big]/dt$, and
$\dot{I}_{y}= \big[I(x_{t},y_{t+dt})-I(x_{t},y_{t})\big]/dt$, with $dt\mapsto 0$.
The rate $\dot{I}_{x}$ can be interpreted as the rate of change of mutual information
between the two subsystems that is due only to the dynamics of X, and similarly
for $\dot{I}_{y}$. One can also represent both $\dot{I}_{x}$ and $\dot{I}_{y}$
by the specific transition rates in both subsystems \cite{horowitz}:

\begin{eqnarray}
  \dot{I}_{x} =
 \sum_{x > x', y} \Big(w^{y}_{xx'}p_{x'y} - w^{y}_{x'x}p_{xy}\Big) \ln\frac{p(y|x)}{p(y|x')},
\end{eqnarray}
and
\begin{eqnarray}
  \dot{I}_{y} =
 \sum_{y > y', x} \Big(w^{x}_{yy'}p_{xy'} - w^{x}_{y'y}p_{xy}\Big) \ln\frac{p(x|y)}{p(x|y')}.
\end{eqnarray}

In the case of sensing the external variable X by the internal variable Y, and with
no feedback from Y to X, the information flows $\dot{I}_{x}$ and $\dot{I}_{y}$
are also called in the literature as nonpredictive information rate (or nostalgia rate)
and learning rate, respectively \cite{still,barato}. One can think about the variable
Y as a molecular sensor or as activity of a neural network learning the dynamics
of the external variable X. Thus the rate of information $\dot{I}_{xy}$ between
internal and external dynamics is the sum of learning rate
($\dot{I}_{y}$ dynamics of Y about external signal X) and the rate of nonpredictive
information ($\dot{I}_{x}$). Using the general bounds derived in Sec. VII on the rate
of mutual information, we can find the bounds on the learning rate, in terms of the
nostalgia rate. Since $|\dot{I}_{x}+\dot{I}_{y}|= |\dot{I}_{xy}| \le B_{I}$, where $B_{I}$
is any of the three bounds on the rate of mutual information (Eqs.(108), (123) and (124)),
we obtain the following general bounds on the learning rate $\dot{I}_{y}$:

\begin{eqnarray}
- B_{I} -  \dot{I}_{x} \le  \dot{I}_{y} \le B_{I} -  \dot{I}_{x}. 
\end{eqnarray}
Thus, the learning rate $\dot{I}_{y}$ about the external signal is bounded
from below and above by the bounds involving temporal Fisher informations
and other kinematic and thermodynamic characteristics.

On the other hand, the nostalgia rate $\dot{I}_{x}$ in this case can be
bounded as \cite{still,ehrich}

\begin{eqnarray}
0 \le  -\dot{I}_{x} \le  \dot{S}(Y|X) - \frac{\dot{Q}_{y}}{k_{B}T}, 
\end{eqnarray}
where the conditional entropy rate
$\dot{S}(Y|X)= \frac{-d}{dt} \sum_{x,y} p_{xy}\ln p(y|x)$, and
$\dot{Q}_{y}$ is the heat flow between the internal system Y and the
thermal environment where
$\dot{Q}_{y}=  -k_{B}T \sum_{y > y', x} \Big(w^{x}_{yy'}p_{xy'} - w^{x}_{y'y}p_{xy}\Big)
\ln\frac{w^{x}_{yy'}}{w^{x}_{y'y}}$.

Combining Eqs. (128) and (129) we obtain the limits on the learning rate
$\dot{I}_{y}$ in terms of the thermodynamic quantities and the derived bounds
$B_{I}$

\begin{eqnarray}
-B_{I} \le  \dot{I}_{y} \le B_{I} + \dot{S}(Y|X) - \frac{\dot{Q}_{y}}{k_{B}T}. 
\end{eqnarray}
This inequality implies that the maximal learning rate of the internal system
about external variable is limited by the internal system entropy production
(which is $\dot{S}(Y|X) - \frac{\dot{Q}_{y}}{k_{B}T}$) and the upper bound
on the rate of MI between the two subsystems.

Eq. (130) can be also used in the context of information erasure and
corresponding heat generation \cite{landauer,still}.
The right hand side of Eq. (130) can be equivalently written as

\begin{eqnarray}
  - \frac{\dot{Q}_{y}}{k_{B}T} \ge  \dot{I}_{er} + (\dot{I}_{y} - B_{I}), 
\end{eqnarray}
where $\dot{I}_{er}= - \dot{S}(Y|X)$ and can be interpreted as the rate of
Landauer erasure of the learned information about X, which is exactly
the rate of decrease of the conditional entropy of Y about X
\cite{still2014}. Inequality (131) is another lower bound on
the heat generation during continuous erase of memory, though not in terms
of nostalgia (as in Eq. 129) but using the learning rate. However, it
should be noted that the bound (131) is less sharp than the bound (129).

\section{\label{sec:level4}Conclusions}

In this work two types of upper bounds on the rates of statistical divergences
were obtained. First type of the bound, with two different inequalities (bounds
B1 and B2 represented by Eqs. (12) and (18)) is very general, independent of the
type of system dynamics, and relates to system's global speed via temporal Fisher
information. Second type, with four different inequalities (bounds B3-B6 represented
by Eqs. (31), (32), and (44)), is less general applying only to Markov systems.
The second type of limitations involves either purely kinematic variables (speeds of
global dynamics and average activities) or a mixture of kinematic and thermodynamic
variables, with the presence of entropy production rate characterizing dissipation
in the system. Generally, the first type of limitations is tighter than the
second, as shown by the numerical example with the one-step Markov process.
The restrictions on the rates of divergences were also used to derive
general bounds on the rates of mutual information between two stochastic variables,
with arbitrary time varying probability distributions. Since statistical divergences
can be thought as generalized information gains \cite{cover}, the present work
suggests a link with information thermodynamics \cite{parrondo,vu}, and it is
also related to recent applications of majorization in thermodynamics
\cite{sagawa}. In particular, this work as well as \cite{sagawa} both provide
a complementary view of out of equilibrium thermodynamic systems in a coherent
manner in terms of information-theoretic and physical variables.
Additionally, the ideas presented here may open new avenues in interdisciplinary
research, e.g., by connecting some areas in computer science, or general computing
either electronic or biological, to the physics of operating algorithms
(e.g. \cite{still,wolpert}). Moreover, the derived bounds on the rates
of mutual information might be useful estimates for information flow in real
neurons \cite{rieke}, artificial deep neural networks \cite{gabrie},
molecular circuits \cite{cheong}, or other systems where exact values are
difficult to obtain \cite{paninski} and require heavy
numerical calculations \cite{kraskov,barato}.
 Finally, it is worth to mention that there are also possible other types of limits
on the rates of statistical divergences, but the goal here was to have bounds
that can be related clearly to the known physical observables.

\section{Acknowledgments}

The work was supported by the Polish National Science Centre (NCN) grant number
2021/41/B/ST3/04300.

\vspace{0.5cm}

\noindent $^{*}$ Email: jkarbowski@mimuw.edu.pl

\appendix

\section{Weakly time dependent exponential distributions applied
  to the bound in Eq. (12).}

The $\alpha$-coefficient and its rate can be computed exactly for the
time dependent exponential distributions. In particular, taking: 
$p_{x}(x)= \nu_{1} e^{-\nu_{1} x}$ and $q_{x}(x)= \nu_{2} e^{-\nu_{2} x}$,
with $\nu_{1}(t)= \nu + \epsilon[\Delta + r_{1}(t)]$, and
$\nu_{2}(t)= \nu + \epsilon r_{2}(t)$, and the small parameter $\epsilon \ll 1$,
yields

\begin{eqnarray}
  C_{\alpha}= \frac{(\nu_{1}/\nu_{2})^{\alpha}}{1 + \alpha[(\nu_{1}/\nu_{2}) -1]}
 \nonumber  \\
 = 1 + \frac{\alpha(\alpha-1)}{2}(\epsilon/\nu)^{2}(r_{1}-r_{2}+\Delta)^{2}
  + O\big(\epsilon^{3}\big),
\end{eqnarray}
and
\begin{eqnarray}
  \frac{dC_{\alpha}}{dt} =
  \alpha(\alpha-1)(\epsilon/\nu)^{2}(r_{1}-r_{2}+\Delta)(\dot{r}_{1}-\dot{r}_{2})
 \nonumber  \\
  + O\big(\epsilon^{3}\big).
\end{eqnarray}
Note that $C_{\alpha}$ is finite provided   $1 + \alpha[(\nu_{1}/\nu_{2}) -1] > 0$.
Otherwise it is infinite.

Temporal Fisher informations are given by

\begin{eqnarray}
 I_{F}(p_{x}) =  \Big( \frac{\epsilon\dot{r}_{1}}{\nu}\Big)^{2}
 + O\big(\epsilon^{3}\big),
 \nonumber  \\
 I_{F}(q_{x}) =  \Big( \frac{\epsilon\dot{r}_{2}}{\nu}\Big)^{2}
 + O\big(\epsilon^{3}\big).
 \end{eqnarray}
Eqs. (A1-A3) allow us to find both sides in Eq. (12) in this case
to the lowest order, as discussed in the main text.

\section{Upper limit on $\langle|\dot{p}/p||(p/q)^{\alpha-1}-C_{\alpha}|^{z}\rangle_{p}$.}
In this Appendix we show how to calculate the bound on
$\langle|\dot{p}/p||(p/q)^{\alpha-1}-C_{\alpha}|^{z}\rangle_{p}$, where the exponent
$z=0$ or $z=1$.

We can generally write

\begin{eqnarray}
  \langle |\dot{p}/p||(p/q)^{\alpha-1}-C_{\alpha}|^{z}\rangle_{p} =
%  \nonumber  \\
  \sum_{n} |\dot{p}_{n}| |(p/q)^{\alpha-1}-C_{\alpha}|^{z}
    \nonumber  \\
 \le  \sum_{nk} |w_{nk}p_{k}-w_{kn}p_{n}| |(p/q)^{\alpha-1}-C_{\alpha}|^{z},
    \nonumber  \\
\end{eqnarray}
where we used master equation in Eq. (22) and the familiar relation
$|x+y| \le |x|+|y|$. Below we restrict the last line in Eq. (B1) in
two different ways: one corresponding to kinematic bound, and second
to mixed kinematic-thermodynamic bound.

\subsubsection{Kinematic bound.}

Let us decompose the last line in Eq. (B1) as

\begin{widetext}
\begin{eqnarray}
  \sum_{nk} |w_{nk}p_{k}-w_{kn}p_{n}| |(p/q)^{\alpha-1}-C_{\alpha}|^{z}
%  \nonumber  \\
  =    \sum_{nk} \sqrt{|w_{nk}p_{k}-w_{kn}p_{n}|}
  \sqrt{|w_{nk}p_{k}-w_{kn}p_{n}|\big[(p/q)^{\alpha-1}-C_{\alpha}\big]^{2z}}
  \nonumber  \\
  \le
   \sum_{nk} \sqrt{(w_{nk}p_{k}+w_{kn}p_{n})}
  \sqrt{(w_{nk}p_{k}+w_{kn}p_{n})\big[(p/q)^{\alpha-1}-C_{\alpha}\big]^{2z}}
% \nonumber  \\
  \le
  \sqrt{ \sum_{nk} (w_{nk}p_{k}+w_{kn}p_{n})}
 \nonumber  \\
\times  \sqrt{ \sum_{nk} (w_{nk}p_{k}+w_{kn}p_{n})\big[(p/q)^{\alpha-1}-C_{\alpha}\big]^{2z}}  
%  \nonumber  \\
= \sqrt{2A_{p}} \sqrt{ \sum_{n} (\dot{p}_{n}+2\overline{w}_{n}p_{n})
    \big[(p/q)^{\alpha-1}-C_{\alpha}\big]^{2z}},    
\end{eqnarray}
\end{widetext}
where we used first the known relation
$|w_{nk}p_{k}-w_{kn}p_{n}| \le (w_{nk}p_{k}+w_{kn}p_{n})$,
and then the Cauchy-Schwartz inequality. In the last line,
we used the master equation (22) for the second factor. Additionally
we introduced two quantities $A_{p}$ and $\overline{w}_{n}$,
which are, respectively, the average activity and the total escape rate
from state $n$. The average activity $A_{p}$ is defined as

\begin{eqnarray}
A_{p}= \frac{1}{2}\sum_{nk}(w_{nk}p_{k}+w_{kn}p_{n}) 
%  \nonumber  \\
  \equiv \sum_{n}\overline{w}_{n}p_{n},
  \nonumber  \\ 
\end{eqnarray}         
from which we also have that the total escape rate from state $n$
is $\overline{w}_{n}= \sum_{k} w_{kn}$.

For $z=0$ the last line in Eq. (B2) simplifies to $2A_{p}$, since
$\sum_{n} \dot{p}_{n}= 0$. This means that

\begin{eqnarray}
\sum_{nk} |w_{nk}p_{k}-w_{kn}p_{n}| \le 2A_{p},
\end{eqnarray}         
which in combination with Eq. (B1) leads to Eq. (24) in the main text.

For $z=1$ we decompose the last factor in the last line of Eq. (B2) into
the sum of two terms: one involving $\dot{p}$ and second $\overline{w}_{n}$,
and then use the Cauchy-Schwartz inequality to both terms. Applying that
procedure yields

\begin{eqnarray}
 \sum_{n}(\dot{p}_{n}+2\overline{w}_{n}p_{n})
    \big[\big(p_{n}/q_{n}\big)^{\alpha-1} - C_{\alpha}\big]^{2}
  \nonumber  \\  
  = \langle (\dot{p}/p)\big[(p/q)^{\alpha-1} - C_{\alpha}\big]^{2}\rangle_{p}
  + 2\langle \overline{w}\big[(p/q)^{\alpha-1} - C_{\alpha}\big]^{2}\rangle_{p},
  \nonumber  \\
  \le
  \Big(\sqrt{I_{F}(p)} + 2\sqrt{\langle\overline{w}^{2}\rangle_{p}}\Big)
   \sqrt{\langle \big[(p/q)^{\alpha-1} - C_{\alpha}\big]^{4}\rangle_{p}},
  \nonumber  \\
\end{eqnarray}
where  $\langle\overline{w}^{2}\rangle_{p}= \sum_{n} \overline{w}_{n}^{2}p_{n}$.
This means that

\begin{eqnarray}
  \sum_{nk} |w_{nk}p_{k}-w_{kn}p_{n}| |(p/q)^{\alpha-1}-C_{\alpha}|
   \nonumber  \\
   \le
 \sqrt{2A_{p}} \Big(\sqrt{I_{F}(p)} + 2\sqrt{\langle\overline{w}^{2}\rangle_{p}}\Big)^{1/2}
   \nonumber  \\
 \langle \big[(p/q)^{\alpha-1} - C_{\alpha}\big]^{4}\rangle_{p}^{1/4},
\end{eqnarray}
which corresponds to Eq. (42) in the main text.

\subsubsection{Kinematic-thermodynamic bound.}

Now we decompose the last line in Eq. (B1) in a different way, which
will allow us to introduce also entropy production rate. 
We have the following sequence of inequalities

\begin{widetext}
\begin{eqnarray}
  \sum_{nk} |w_{nk}p_{k}-w_{kn}p_{n}| |\big(p_{n}/q_{n}\big)^{\alpha-1} - C_{\alpha}|^{z}
%  \nonumber  \\
  = \sum_{nk} \sqrt{(w_{nk}p_{k}-w_{kn}p_{n})
    \ln\frac{w_{nk}p_{k}}{w_{kn}p_{n}}}
  \sqrt{\frac{(w_{nk}p_{k}-w_{kn}p_{n})}
    {\ln\frac{w_{nk}p_{k}}{w_{kn}p_{n}}}}
  \nonumber  \\
 \times |\big(p_{n}/q_{n}\big)^{\alpha-1} - C_{\alpha}|^{z}
  \le  \sqrt{\sum_{nk}(w_{nk}p_{k}-w_{kn}p_{n})
    \ln\frac{w_{nk}p_{k}}{w_{kn}p_{n}}}
  \sqrt{\sum_{nk}\frac{(w_{nk}p_{k}-w_{kn}p_{n})}
    {\ln(w_{nk}p_{k}/w_{kn}p_{n})} \big[\big(p_{n}/q_{n}\big)^{\alpha-1} - C_{\alpha}\big]^{2z}}
  \nonumber  \\
\end{eqnarray}
\end{widetext}
where we used in the second line the Cauchy-Schwartz inequality.
The first factor in the last line of Eq. (B7) is the coarse-grained entropy
production rate $\dot{S}_{p}$, i.e. \cite{schnakenberg,maes1,esposito}

\begin{eqnarray}
  \dot{S}_{p}=  \frac{1}{2}
  \sum_{nk}(w_{nk}p_{k}-w_{kn}p_{n})
  \ln\frac{w_{nk}p_{k}}{w_{kn}p_{n}}.
\end{eqnarray}
It is worth to mention that coarse-grained entropy production
$\dot{S}_{p}$ satisfies inequality
$\dot{S}_{p} \le  \frac{1}{2}
  \sum_{s}\sum_{nk} \big(w_{nk}^{(s)}p_{k}-w_{kn}^{(s)}p_{n}\big)
  \ln\big(w_{nk}^{(s)}p_{k}/w_{kn}^{(s)}p_{n}\big)$,
  which means that $\dot{S}_{p}$ is the lower bound on the true physical
  entropy production caused by distinct microscopic processes \cite{esposito}.
However, in our case this is a sufficient estimation.

The second factor on the right in the last line of Eq. (B7) can be bounded by
the logarithmic mean, as in Eq. (5):

\begin{eqnarray}
\frac{(w_{nk}p_{k}-w_{kn}p_{n})}
      {\ln(w_{nk}p_{k})-\ln(w_{kn}p_{n})}
 \le  \frac{1}{2}(w_{nk}p_{k}+w_{kn}p_{n}).
  \nonumber  \\ 
\end{eqnarray}         
Combining Eqs. (B7-B9) we obtain the following inequality

\begin{eqnarray}
  \sum_{nk} |w_{nk}p_{k}-w_{kn}p_{n}| |\big(p_{n}/q_{n}\big)^{\alpha-1} - C_{\alpha}|^{z}
 \nonumber  \\
  \le \sqrt{2\dot{S}_{p}}
 \sqrt{\sum_{nk} \frac{1}{2}(w_{nk}p_{k}+w_{kn}p_{n})
   \big[\big(p_{n}/q_{n}\big)^{\alpha-1} - C_{\alpha}\big]^{2z}}
 \nonumber  \\
= \sqrt{\dot{S}_{p}} \sqrt{\sum_{n}(\dot{p}_{n}+2\overline{w}_{n}p_{n})
   \big[\big(p_{n}/q_{n}\big)^{\alpha-1} - C_{\alpha}\big]^{2z}}.
  \nonumber  \\
\end{eqnarray}

For $z=0$ this inequality implies:

\begin{eqnarray}
\sum_{nk} |w_{nk}p_{k}-w_{kn}p_{n}| \le \sqrt{2\dot{S}_{p}}\sqrt{A_{p}},
\end{eqnarray}         
which in combination with Eq. (B1) produces Eq. (26) in the main text.

For $z=1$, after applying Eq. (B5), Eq. (B10) gives

\begin{eqnarray}
  \sum_{nk} |w_{nk}p_{k}-w_{kn}p_{n}| |(p/q)^{\alpha-1}-C_{\alpha}|
   \nonumber  \\
   \le
   \sqrt{\dot{S}_{p}}
   \Big(\sqrt{I_{F}(p)} + 2\sqrt{\langle\overline{w}^{2}\rangle_{p}}\Big)^{1/2}
   \nonumber  \\
 \langle \big[(p/q)^{\alpha-1} - C_{\alpha}\big]^{4}\rangle_{p}^{1/4},
\end{eqnarray}
which corresponds to Eq. (42) in the main text.

%\vspace{0.5cm}

\section{Upper limit on $\langle |X-\langle X\rangle|^{3}\rangle$
  and related inequalities.}
In this Appendix we provide three different bounds on
$\langle |X-\langle X\rangle|^{3}\rangle$, where $X$ is some non-negative
random variable, for which first four moments exist. (We require non-negativity
of $X$, since we deal with such cases in this paper.)

The first method generates the following inequality

\begin{eqnarray}
  \langle |X-\langle X\rangle|^{3}\rangle
  \le \langle X^{3}\rangle - \langle X\rangle \langle X^{2}\rangle,
\end{eqnarray}  
which can be justifying in few step, as

$\langle |X-\langle X\rangle|^{3}\rangle =
\langle (X-\langle X\rangle)^{2}|X-\langle X\rangle|\rangle \le
\langle (X-\langle X\rangle)^{2}(X+\langle X\rangle)\rangle =
\langle X\rangle\langle (X-\langle X\rangle)^{2}\rangle +
\langle X(X-\langle X\rangle)^{2}\rangle $
\\
and by performing the averages. In the above a simple inequality
valid for $X \ge 0$ was used, namely,
$|X-\langle X\rangle| \le X + \langle X\rangle$.

The second method generates more complicated inequality
  
\begin{eqnarray}
  \langle |X-\langle X\rangle|^{3}\rangle
  \le \sqrt{\langle X^{2}\rangle - \langle X\rangle^{2}}  
 \nonumber \\
\times  \sqrt{\langle X^{4}\rangle + 6\langle X\rangle^{2}\langle X^{2}\rangle
  - 4\langle X\rangle\langle X^{3}\rangle - 3\langle X\rangle^{4}},
 \nonumber \\
\end{eqnarray}  
which follows from applying the Cauchy-Swartz inequality as

\begin{eqnarray*}
\langle |X-\langle X\rangle|^{3}\rangle =
\langle (X-\langle X\rangle)^{2}|X-\langle X\rangle|\rangle
 \nonumber \\
\le \sqrt{\langle (X-\langle X\rangle)^{4}\rangle}
\sqrt{\langle (X-\langle X\rangle)^{2}\rangle},
\end{eqnarray*}  
and executing some algebra under the square roots.

The third method uses Minkowski inequality, known in generality
as: $\langle |X+Y|^{s}\rangle^{1/s} \le
\langle |X|^{s}\rangle^{1/s} +  \langle |Y|^{s}\rangle^{1/s}$.
In our case we obtain

\begin{eqnarray}
  \langle |X-\langle X\rangle|^{3}\rangle
  \le \Big(\langle X^{3}\rangle^{1/3} + \langle X\rangle\Big)^{3}.
\end{eqnarray}

It clear that the upper bound in Eq. (C3) is larger than $\langle X^{3}\rangle$,
and hence larger than the bound in Eq. (C1). This implies that Eq. (C1) provides
a tighter bound than Eq. (C3). Moreover, the formula (C1) is simpler than the
formula (C2), and it requires lower moments. For these reasons we use Eq. (C1)
for estimations in this paper. For example, Eq. (28) can be obtained by
substituting $(p/q)^{\alpha-1}$ for $X$, and noting that
$C_{\alpha}= \langle (p/q)^{\alpha-1}\rangle_{p}$.

Now let us find the upper limit on a related average, which appears in Eq. (39),
i.e., $\langle |\ln(X)-\mu|^{3}\rangle$, where $\mu= \langle \ln(X)\rangle$,
and $\mu$ can be negative.
We have $\langle |\ln(X)-\mu|^{3}\rangle= \langle |\ln(X/e^{\mu})|^{3}\rangle$.
From Eq. (5) for arbitrary positive numbers $x, y$ we have

\begin{eqnarray}
|\ln(x/y)| \le |\sqrt{x/y} - \sqrt{y/x}|,
\end{eqnarray}
which implies for $x= X$ and $y= e^{\mu}$ the following series of inequalities

\begin{eqnarray}
  \langle |\ln(X)-\mu|^{3}\rangle
  \le \langle |\sqrt{X/e^{\mu}} - \sqrt{e^{\mu}/X}|^{3}\rangle
  \nonumber \\
=  \langle \big[\sqrt{X/e^{\mu}} - \sqrt{e^{\mu}/X}\big]^{2}
|\sqrt{X/e^{\mu}} - \sqrt{e^{\mu}/X}|\rangle
  \nonumber \\
  \le  \langle \big[\sqrt{X/e^{\mu}} - \sqrt{e^{\mu}/X}\big]^{2}
  \big(\sqrt{X/e^{\mu}} + \sqrt{e^{\mu}/X}\big)\rangle
 \nonumber \\
 = e^{-3\mu/2}\langle X^{3/2}\rangle - e^{-\mu/2}\langle X^{1/2}\rangle
 \nonumber \\
 - e^{\mu/2}\langle X^{-1/2}\rangle + e^{3\mu/2}\langle X^{-3/2}\rangle.
\end{eqnarray}
Similarly, we can estimate $\langle [\ln(X)-\mu]^{4}\rangle$, which appears
in Eq. (43):

\begin{eqnarray}
  \langle [\ln(X)-\mu]^{4}\rangle
  \le \langle [\sqrt{X/e^{\mu}} - \sqrt{e^{\mu}/X}]^{4}\rangle
  \nonumber \\
 = e^{-2\mu}\langle X^{2}\rangle - 4e^{-\mu}\langle X\rangle + 6
 - 4e^{\mu}\langle X^{-1}\rangle + e^{2\mu}\langle X^{-2}\rangle.
 \nonumber \\
\end{eqnarray}

Obviously, the inequalities (C5) and (C6) require that several first few
fractional moments (even of negative order) exist.

\section{Mutual information and temporal Fisher information for the
  model in Eq. (114).}

Mutual information $I(x,y)$ between $x$ and $y$ variables is
$I(x,y)= \langle \ln\big( \frac{p_{xy}}{p_{x}p_{y}} \big)\rangle$, which
for the model in Eq. (114) translates to

\begin{eqnarray}
  I(x,y)= \ln\big(1 + \frac{\epsilon r(t)}{\kappa_{1}\kappa_{2}}\big)
  + \langle \ln\big(1 + \epsilon r(t)xy\big)\rangle
  \nonumber \\
  - \langle \ln\big(1 + \epsilon r(t)x/\kappa_{2}\big)\rangle
  - \langle \ln\big(1 + \epsilon r(t)y/\kappa_{1}\big)\rangle.
\end{eqnarray}
Next, we approximate the logarithms to the second order in $\epsilon$,
according to the formula: $\ln(1+x)= x - x^{2}/2 + O(x^{3})$. After
that step the mutual information reads:

\begin{eqnarray}
  I(x,y)= \epsilon r(t)\Big( \Big[ \frac{1}{\kappa_{1}\kappa_{2}} +
  \langle xy\rangle - \frac{1}{\kappa_{2}}\langle x\rangle 
  - \frac{1}{\kappa_{1}}\langle y\rangle \Big]
  \nonumber \\
  + \frac{1}{2}\epsilon r(t)\Big[
    \frac{1}{\kappa_{2}^{2}}\langle x^{2}\rangle
  +  \frac{1}{\kappa_{1}^{2}}\langle y^{2}\rangle
  -  \langle x^{2}y^{2}\rangle
  -  \frac{1}{(\kappa_{1}\kappa_{2})^{2}} \Big] \Big) + O(\epsilon^{3}).
  \nonumber \\
\end{eqnarray}
The averages in Eq. (D2) are given to the lowest order by

\begin{eqnarray}
  \langle x\rangle =  \frac{1}{\kappa_{1}}
  \Big( 1+ \frac{\epsilon r(t)}{\kappa_{1}\kappa_{2}}\Big) + O(\epsilon^{2}),
\end{eqnarray}
\begin{eqnarray}
  \langle y\rangle =  \frac{1}{\kappa_{2}}
  \Big( 1+ \frac{\epsilon r(t)}{\kappa_{1}\kappa_{2}}\Big) + O(\epsilon^{2}),
\end{eqnarray}
 \begin{eqnarray}
 \langle xy\rangle =  \frac{1}{\kappa_{1}\kappa_{2}}
  \Big( 1+ 3\frac{\epsilon r(t)}{\kappa_{1}\kappa_{2}}\Big) + O(\epsilon^{2}),
\end{eqnarray}
 \begin{eqnarray}
  \langle x^{2}\rangle =  \frac{2}{\kappa_{1}^{2}}
  \Big( 1+ 5\frac{\epsilon r(t)}{\kappa_{1}\kappa_{2}}\Big) + O(\epsilon^{2}),
 \end{eqnarray}
 \begin{eqnarray}
  \langle y^{2}\rangle =  \frac{2}{\kappa_{2}^{2}}
  \Big( 1+ 5\frac{\epsilon r(t)}{\kappa_{1}\kappa_{2}}\Big) + O(\epsilon^{2}),
 \end{eqnarray}
 \begin{eqnarray} 
  \langle x^{2}y^{2}\rangle =  \frac{4}{(\kappa_{1}\kappa_{2})^{2}}
  \Big( 1+ 8\frac{\epsilon r(t)}{\kappa_{1}\kappa_{2}}\Big) + O(\epsilon^{2}).
%  \nonumber \\
\end{eqnarray}
Insertion of the averages in Eqs. (D3-D8) to Eq. (D2) with some algebra produces
Eq. (116) in the main text.

Temporal Fisher information $I_{F}(p_{x})= \langle (\dot{p}_{x}/p_{x})^{2}\rangle$
can be obtained by finding the ratio $\dot{p}_{x}/p_{x}$, which is

\begin{eqnarray}
 \frac{\dot{p}_{x}}{p_{x}}=
\epsilon \frac{\dot{r}(t)}{\kappa_{2}}
\Big[  \frac{1}{\kappa_{1}}\big(-1 + \epsilon\frac{r(t)}{\kappa_{1}\kappa_{2}}\big)
 + x \big(1 - \epsilon\frac{r(t)x}{\kappa_{2}}\big) \Big] 
+ O(\epsilon^{3}).
  \nonumber \\
\end{eqnarray}
Averaging the square of this expression yields Eq. (117) in the main text.
In the similar way one can evaluate temporal Fisher information for the joint
probability density $p_{xy}$.

%\bibliography{apssamp} % Produces the bibliography via BibTeX.

\begin{thebibliography}{99}

\bibitem{renyi}
  A. Renyi. {\it On measures of entropy and information.} Proc. 4th Berkeley Symposium
  on Mathematics, Statistics and Probability, pp. 547-561 (1960). Univ. of California
  Press: Berkeley, CA.
  
\bibitem{csiszar}
  I. Csiszar. Information-type measures of difference of probability distributions
  and indirect observations. {\it Studia Scientiarum Mathematicarum Hungarica}
  {\bf 2}, 299-318 (1967).

\bibitem{tsallis1}
  C. Tsallis. Generalized entropy-based criterion for consistent testing.
  {\it Phys. Rev.} E  {\bf 58}, 1442-1445 (1998).

\bibitem{kullback}
  S. Kullback, R.A. Leibler. On information and sufficiency.
  {\it Ann. Math. Stat.} {\bf 22}, 79-86 (1951).
  
\bibitem{cichocki}
  A. Cichocki, S-I. Amari. Families of alpha- beta- and gamma-divergences:
  flexible and robust measures of similarity. {\it Entropy} {\bf 12},
  1532-1568 (2010).

\bibitem{hatano}  
  T. Hatano, S. Sasa. Steady-state thermodynamics of Langevin systems.
  {\it Phys. Rev. Lett.}  {\bf 86}, 3463-3466 (2001).

\bibitem{esposito}  
  M. Esposito, C. Van den Broeck. Three faces of the second law. I. Master
  equation formulation. {\it Phys. Rev.} E {\bf 82}, 011143 (2010).

\bibitem{still}  
  S. Still, D.A. Sivak, A.J. Bell, G.E. Crooks. Thermodynamics of
  prediction.  {\it Phys. Rev. Lett.}  {\bf 109}, 120604 (2012).
  
\bibitem{falasco}  
  G. Falasco, M. Esposito, J.-C. Delvenne. Beyond thermodynamic 
  uncertainty relations: nonlinear response, error-dissipation
  trade-offs, and speed limits.  {\it J. Phys.} A {\bf 55}, 124002 (2022).

  
\bibitem{cover}
T.M. Cover, J.A. Thomas. {\it Elements of Information Theory}. Wiley and Sons (2006).

\bibitem{reid}
  M.D. Reid, R.C. Williamson. Information, divergence and risk for binary experiments.
  {\it J. Mach. Learn. Res.} {\bf 12}, 731-817 (2011).
  
\bibitem{liese}
F. Liese, I. Vajda. On divergences and informations in statistics and information theory.
  {\it IEEE Trans. Inform. Theory} {\bf 52}, 4394-4412 (2006). 

\bibitem{amari}
S-I. Amari, H. Nagaoka. {\it Methods of Information Geometry}. Oxford Univ. Press (2000).

\bibitem{sason}
  I. Sason, S. Verdu. f-Divergence inequalities. {\it IEEE Trans. Inf. Theory}
  {\bf 62}, 5973-6006 (2016).

\bibitem{morimoto}
  T. Morimoto. Markov processes and the H-theorem. {\it J. Phys. Soc. Jap.}
  {\bf 12}, 328-331 (1963).
  
\bibitem{risken}
H. Risken. {\it The Fokker-Planck Equation}. Springer, Berlin (1989).

\bibitem{borland}
  L. Borland, A.R. Plastino, C. Tsallis. Information gain within nonextensive
  thermostatistics.  {\it J. Math. Phys.} {\bf 39}, 6490-6501 (1998).  

\bibitem{yamano2}
  T. Yamano. de Bruijn-type identity for systems with flux.
  {\it Eur. Phys. J.} B {\bf 86}, 363 (2013).  

\bibitem{gorban}
  A.N. Gorban. General H-theorem and entropies that violate the second law.
  {\it Entropy} {\bf 16}, 2408-2432 (2014).

\bibitem{crooks}
  G.E. Crooks. Measuring thermodynamic length.  {\it Phys. Rev. Lett.}
  {\bf 99}, 100602 (2007).

\bibitem{parrondo}
  J.M. Parrondo, J.M. Horowitz, T. Sagawa. Thermodynamics of information.
  {\it Nat. Phys.} {\bf 11}, 131 (2015).
    
\bibitem{vu}
  T.V. Vu, K. Saito. Thermodynamic unification of optimal transport:
  Thermodynamic uncertainty relation, minimum dissipation, and
  thermodynamic speed limits. {\it Phys. Rev.} X
  {\bf 13}, 011013 (2023).


% thermodynamic-kinematic inequalities
\bibitem{ito}
  S. Ito, A. Dechant. Stochastic time evolution, information geometry
  and the Cramer-Rao bound. {\it Phys. Rev. X} {\bf 10}, 021056 (2020).

\bibitem{nicholson}
  S.B. Nicholson, L.P. Garcia-Pintos, A. del Campo, J.R. Green.
  Time-information uncertainty relations in thermodynamics.
  {\it Nature Phys.} {\bf 16}, 1211-1215 (2020).

\bibitem{shiraishi}
  N. Shiraishi, K. Funo, K. Saito. Speed limit for classical stochastic
  processes.  {\it Phys. Rev. Lett.} {\bf 121}, 070601 (2018).

\bibitem{koyuk}
  T. Koyuk, U. Seifert. Thermodynamic uncertainty relations for time-dependent
  driving. {\it Phys. Rev. Lett.} {\bf 125}, 260604 (2020).
  
\bibitem{vo}
V.T. Vo, T.V. Vu, Y. Hasegawa. Unified thermodynamic-kinetic uncertainty
relation.  {\it J. Phys. } A {\bf 55}, 405004 (2022).
% end thermodynamic-kinematic inequalities

\bibitem{kolchinsky}
  A. Kolchinsky, B.D. Tracey. Estimating mixture entropy with pairwise
  distances.   {\it Entropy} {\bf 19}, 361 (2017).


  
\bibitem{finner}
  H. Finner. A generalization of H\"older's inequality and some probability
  inequalities. {\it Annals of Probability}  {\bf 20}, 1893-1901 (1992).

\bibitem{carlson}
  B.C. Carlson. Some inequalities for hypergeometric functions.
  {\it Proc. Amer. Math. Soc.} {\bf 17}, 32-39 (1966).

% temporal fisher information:
\bibitem{frieden}
B.R. Frieden. {\it Science from Fisher Information: A Unification,} 
2nd ed.  Cambridge Univ. Press, Cambridge, UK (2004).
  

  
% relative fisher information:
\bibitem{otto}
  F. Otto, C. Villani. Generalization of an inequality by Talagrand and links
  with the logarithmic Sobolev inequality.
  {\it J. Funct. Analysis} {\bf 173}, 361-400 (2000).

\bibitem{yamano1}
  T. Yamano. Phase space gradient of dissipated work and information:
  A role of relative Fisher information.
  {\it J. Math. Phys.} {\bf 54}, 113301 (2013).

%  end relative fisher

\bibitem{schnakenberg}
  J. Schnakenberg. Network theory of microscopic and macroscopic behavior
  of master equation systems. {\it Rev. Mod. Phys.} {\bf 48}, 571 (1976).

\bibitem{baiesi}
  M. Baiesi, C. Maes, B. Wynants. Fluctuations and response of nonequilibrium
  states.  {\it Phys. Rev. Lett.} {\bf 103}, 010602 (2009).

  
\bibitem{maes1}  
  C. Maes, K. Netocny. Time-reversal and entropy. {\it J. Stat. Phys.}
  {\bf 110}, 269-310 (2003).
  
\bibitem{maes2}  
 C. Maes. Frenesy: Time-symmetric dynamical activity in nonequilibria.
  {\it Physics Reports} {\bf 850}, 1-33 (2020).

  
\bibitem{vankampen}
Van Kampen NG (2007) {\it Stochastic Processes in Physics and Chemistry}. Elsevier:
Amsterdam.

\bibitem{margolus}
N. Margolus, L.B. Levitin. The maximum speed of dynamical evolution.
  {\it Physica} D {\bf 120}, 188 (1998).

\bibitem{shanahan}
  B. Shanahan, A. Chenu, N. Margolus, A. del Campo. Quantum speed limits across
  the quantum-to-classical transition.  {\it Phys. Rev. Lett.} {\bf 120}, 070401 (2018).

\bibitem{hamazaki}
  R. Hamazaki. Speed limits for macroscopic transitions.
  {\it PRX Quantum} {\bf 3}, 020319 (2022).
 
\bibitem{garcia}
  L.P. Garcia-Pintos, S.B. Nicholson, J.R. Green, A. del Campo, A.V. Gorshkov.
 Unifying quantum and classical speed limits on observables.
  {\it Phys. Rev. X} {\bf 12}, 011038 (2022).
 
\bibitem{deffner}
  S. Deffner, S. Campbell. Quantum speed limits: From Heisenberg's uncertainty
  principle to optimal quantum control. {\it J. Phys.} A {\bf 50}, 453001 (2017).
 

\bibitem{barato2015}
  A.C. Barato, U. Seifert. Thermodynamic uncertainty relation for biomolecular
  processes.  {\it Phys. Rev. Lett.} {\bf 114}, 158101 (2015).

\bibitem{li}
  J. Li, J.M. Horowitz, T.R. Gingrich, N. Fakhri. Quantifying dissipation
  using fluctuating currents.  {\it Nat. Commun.} {\bf 10}, 1666 (2019).

\bibitem{vu2020}
  T. Van Vu, V.T. Vo, Y. Hasegawa. Entropy production estimation with optimal
  current.  {\it Phys. Rev.} E {\bf 101}, 042138 (2020).

\bibitem{skinner}
  D.J. Skinner, J. Dunkel. Estimating entropy production from waiting
  time distributions.  {\it Phys. Rev. Lett.} {\bf 127}, 198101 (2021).

\bibitem{salazar}
  D.S.P. Salazar. Lower bound for entropy production rate in stochastic
  systems far from equilibrium.  {\it Phys. Rev.} E {\bf 106}, L032101 (2022).

\bibitem{dechant}
  A. Dechant. Minimum entropy production, detailed balance and Wasserstein
  distance for continuous-time Markov processes.  {\it J. Phys.} A {\bf 55},
  094001 (2022).

\bibitem{esposito2011}
  M. Esposito, C. van den Broeck. Second law and Landauer principle far
  from equilibrium. {\it EPL} {\bf 95}, 40004 (2011).

\bibitem{novikov}
  E.A. Novikov. Functionals and the random-force method in turbulence
  theory. {\it Soviet Physics JETP} {\bf 20}, 1290-1294 (1965).

\bibitem{miller}
  P. Miller, A.M. Zhabotinsky, J.E. Lisman, X-J. Wang. The stability
  of stochastic CaMKII switch: dependence on the number of enzyme
  molecules and protein turnover. {\it PLoS Biol.} {\bf 3},
  e107 (2005).
  
\bibitem{petersen}
  C.C. Petersen, R.C. Malenka, R.A. Nicoll, J.J. Hopfield. All-or-none
  potentiation at CA3-CA1 synapses.  {\it Proc. Natl. Acad. USA} {\bf 95},
  4732-4737 (1998).

\bibitem{karbowski2019}
J. Karbowski. Metabolic constraints on synaptic learning and memory.
  {\it J. Neurophysiol.} {\bf 122}, 1473-1490 (2019).
  
\bibitem{dayan}
P. Dayan, L.F. Abbott. {\it Theoretical Neuroscience}. MIT Press (2000).

\bibitem{chklovskii}
D.B. Chklovskii, B.W. Mel, K. Svoboda. Cortical rewiring and information
  storage. {\it Nature} {\bf 431}, 782-788 (2004).

\bibitem{kasai}
H. Kasai, N.E. Ziv, H. Okazaki, S. Yagishita, T. Toyoizumi.
Spine dynamics in the brain, mental disorders, and artificial neural networks. 
  {\it Nat. Rev. Neurosci.} {\bf 22}, 407-422 (2021).


\bibitem{karbowski2023}
J. Karbowski, P. Urban. Information encoded in volumes and areas of dendritic
  spines is nearly maximal across mammalian brains.
  {\it Sci. Rep.} {\bf 13}, 22207 (2023).

\bibitem{karbowski2024}
J. Karbowski, P. Urban. Cooperativity, information gain, and energy cost
 during early LTP in dendritic spines. {\it Neural Comput.} {\bf 36}, 271-311 (2024).

\bibitem{bialek2001}
  W. Bialek, I. Nemenman, N. Tishby. Predictability, complexity, and learning.
  {\it Neural Comput.} {\bf 13}, 2409-2463 (2001).

\bibitem{lang}
  A.H. Lang, C.K. Fisher, T. Mora, P. Mehta. Thermodynamics of statistical
  inference by cells.  {\it Phys. Rev. Lett.} {\bf 113}, 148103 (2014).

\bibitem{palmer}
  S.E. Palmer, O. Marre, M.J. Berry, W. Bialek. Predictive information in a
  sensory population.  {\it Proc. Natl. Acad. USA} {\bf 112},
  6908-6913 (2015).

\bibitem{still2014}
S. Still. Information bottleneck approach to predictive inference.
  {\it Entropy} {\bf 16}, 968-989 (2014).


 % bipartite systems
\bibitem{hartich}
  D. Hartich, A.C. Barato, U. Seifert. Stochastic thermodynamics of
  bipartite systems: transfer entropy inequalities and a Maxwell's
  demon interpretation. {\it J. Stat. Mech.} P02016 (2014).
  
\bibitem{diana}
  G. Diana, M. Esposito. Mutual entropy production in
  bipartite systems. {\it J. Stat. Mech.} P04010 (2014).

\bibitem{horowitz}
  J.M. Horowitz, M. Esposito. Thermodynamics with continuous information
  flow. {\it Phys. Rev. X} {\bf 4}, 031015 (2014).


\bibitem{barato}
  A.C. Barato, D. Hartich, U. Seifert. Information-theoretic
  versus thermodynamic entropy production in autonomous
  sensory networks. {\it Phys. Rev.} E {\bf 87}, 042104 (2013).

\bibitem{paninski}
L. Paninski. Estimation of entropy and mutual information. {\it Neural Comput.}
{\bf 15}, 1191-1253.   

% numerical bound on MI rate
\bibitem{kraskov}
  A. Kraskov, H. Stogbauer, P. Grassberger. Estimating mutual information
  {\it Phys. Rev.} E {\bf 69}, 066138 (2004).

\bibitem{allahverdyan}
  A.E. Allahverdyan, D. Janzing, G. Mahler. Thermodynamic efficiency of information
  and heat flow. {\it J. Stat. Mech.} P09011 (2009).

\bibitem{ehrich}
J. Ehrich, D.A. Sivak. Energy and information flows in autonomous systems.
  {\it Front. Phys.} {\bf 11}: 1108357 (2023).

\bibitem{landauer}
  R. Landauer. Irreversibility and heat generation in the computing
  process. {\it IBM J. Res. Develop.} {\bf 5}, 183-191 (1961).
  

\bibitem{sagawa}
  T. Sagawa. {\it Entropy, divergence, and majorization in classical
    and quantum thermodynamics.}
Singapore: Springer (2022). 
  
\bibitem{wolpert}
D.H. Wolpert. The stochastic thermodynamics of computation. {\it J. Phys.} A
 {\bf 52}, 193001 (2019).
  
\bibitem{rieke}
F. Rieke, D. Warland, R. de Ruyter, W. Bialek.
{\it Spikes: Exploring the neural code.}
Cambridge, MA: MIT Press (1999). 
  
\bibitem{gabrie}
M. Gabrie, A. Manoel, C. Luneau, et al. Entropy and mutual information in
 models of deep neural networks. {\it J. Stat. Mech.} 124014 (2019).

\bibitem{cheong}
R. Cheong, A. Rhee, C.J. Wang, I. Nemenman, A. Levchenko. Information transduction
capacity of noisy biochemical signaling networks. {\it Science} {\bf 334}, 354-358
  (2011).


 

\end{thebibliography}

\end{document}